\documentclass[prd,nofootinbib,superscriptaddress,preprint]{revtex4}
\usepackage[T1]{fontenc}
\usepackage{amsmath,amssymb}
\usepackage{epsfig}
\usepackage{graphicx,array}
\usepackage[usenames,dvipsnames]{color}
\usepackage{slashed}
\usepackage{comment}
\usepackage[colorlinks,citecolor=blue]{hyperref}
\usepackage{pdfpages}
\usepackage{color}
\usepackage{subfigure}
\begin{document}

\title{Matter-antimatter asymmetry in minimal inverse seesaw framework with $A_4$ modular symmetry} 
	\author{Gourab Pathak}
	\email{gourabpathak7@gmail.com}
	\affiliation{Department of Physics, Tezpur University, Napaam, Assam, India-784028}

\author{Mrinal Kumar Das}
\email{mkdas@tezu.ernet.in}
\affiliation{Department of Physics, Tezpur University, Napaam, Assam, India-784028}

\begin{abstract}
We propose a minimal inverse seesaw framework based on $A_4$ modular symmetry. We have studied the neutrino oscillation parameters in our work and our model excludes some $3 \sigma$ values of the mixing angle $\theta_{23}$. Also, there is a clear linear relation between the mixing angles $\theta_{12}$ and $\theta_{23}$ found in the allowed $3 \sigma$ region. We also examine whether the parameter points consistent with neutrino oscillation data simultaneously comply with the experimental limits on lepton flavor violating (LFV) decays, specifically: $\mu \longrightarrow e \gamma$, $\tau \longrightarrow e \gamma$, and $\tau \longrightarrow \mu \gamma$. We have also investigated the matter-antimatter asymmetry of our universe via the resonant leptogenesis mechanism. Here, we present the contribution of lepton number conserving scattering processes mediated by the \( Z' \) boson in the context of leptogenesis.
\end{abstract}
%\end{titlepage}

\maketitle 

%%%%%%%%%%%%%%%%%%% Intro %%%%%%%%%%%%%%%%%%%%
\section{Introduction}\label{sec:intro}
One of the outstanding problems in particle physics is the flavor structure of the Standard Model (SM) fermions. In the SM, the mass hierarchy between the three generations of charged fermions in both the quark and the lepton sectors is not addressed by gauge symmetry and we expect it can be explained in the Beyond Standard Model framework (BSM).
The peculiar mixing between the leptons and CP violation in neutrino oscillation could be understood via discrete non-abelian symmetry. But, this approach needs a substantial number of flavon fields to explain the observed mixing. Recently, a modular flavor symmetry has been proposed to study the flavor structure \cite{Feruglio:2017spp}. In this framework, the Yukawa couplings of the SM take modular form and they are function of the complex field \textit{modulus} $\tau$, the VEV of which breaks the modular group completely. The flavor structure can be achieved without requiring many flavon fields in this framework. The modulus $\tau$ describes the geometry of compactified extra-dimension in superstring theory \cite{Lauer:1989ax,Lauer:1990tm}. The finite modular group $\Gamma_{N}$ of level $N\in \mathbb{Z}$ is isomorphic to the more familiar non-abelian groups $S_3$, $A_4$, $S_4$ and $A_5$. Currently, a substantial number of modular symmetric models are available in the literature including $\Gamma_{2} \cong S_{3}$ \cite{Meloni:2023aru,Okada:2019xqk}, $\Gamma_{3} \cong A_{4}$ \cite{Kashav:2021zir,Kashav:2022kpk,Singh:2024imk,Nomura:2019jxj,Nomura:2019xsb,Behera:2020lpd,Abbas:2020qzc,Altarelli:2005yx}, $\Gamma_{4} \cong S_{4}$ \cite{Kobayashi:2019xvz, Penedo:2018nmg, Wang:2019ovr, Zhang:2021olk}, $\Gamma_{5} \cong A_{5}$ \cite{Novichkov:2018nkm,Ding:2019xna,Behera:2021eut,Behera:2022wco} and also double cover modular $\Gamma_{N}'$ of finite group such as $A_{4}$ \cite{Mishra:2023cjc,Okada:2022kee}, $A_{5}$ \cite{Wang:2020lxk} etc.
\par
In the SM, neutrinos are massless due to the absence of counterpart of left-handed neutrinos. Neutrino oscillation observations confirm that neutrinos are indeed massive and their small mass can be explained via $see-saw$ mechanism. To achieve this, we need to expand the SM particle content, and depending upon the BSM particle and their representation, various types of see-saw mechanism have been proposed, e.g. Type-I \cite{Brdar:2019iem,Schechter:1981cv}, Type-II \cite{Barrie:2022cub,SanchezVillamizar:2019fce,Rodejohann:2004cg}, Type-III \cite{Albright:2003xb,Ashanujjaman:2021jhi,Biswas:2019ygr}, etc. But, these mechanism needs heavy BSM particles, which are difficult to search in present ongoing experiments. The $inverse-seesaw$ \cite{Dias:2012xp,Nomura:2019xsb,Zhang:2021olk,Chakraborty:2021azg,Gogoi:2023jzl,Borah:2017dmk} mechanism is one of the popular seesaw mechanisms for generating tiny neutrino mass as it needs TeV scale BSM heavy particles, which can be detected in current experiments.
\par
Besides the aforementioned problems, matter-antimatter asymmetry of our universe is a longstanding puzzle, which cannot be solved in the SM framework. The present value of this asymmetry, also known as Baryon Asymmetry of Universe (BAU), is given by \cite{Planck:2018vyg}:
\begin{equation}
    \eta_{B}=\frac{n_{B}-n_{\bar{B}}}{n_{\gamma}}=(6.12\pm0.04)\times10^{-10}.
\end{equation}
where $\eta_{B}$ is baryon asymmetry per unit photon density.
The three Sakharov conditions need to be satisfied for dynamical generation of baryon asymmetry \cite{Sakharov:1967dj}:
\begin{enumerate}
    \item Baryon number (B) violation.
    \item C or CP-violation.
    \item Out-of-equillibrium condition.
\end{enumerate}
The complex modulus $\tau$ is the source of CP-violation and out of equilibrium decay of heavy neutrinos produce the lepton asymmetry, which can be transformed into baryon asymmetry via the SM sphaleron process. This mechanism is known as \textit{leptogenesis} \cite{Davidson:2008bu,Buchmuller:2004nz,Buchmuller:2005eh,Pilaftsis:1997jf,Blanchet:2010kw,Plumacher:1996kc,Barbieri:1999ma,FileviezPerez:2021hbc,Flanz:1996fb}. If the mass-splitting of heavy neutrinos becomes comparable to their decay width, then BAU is achieved by the low scale $resonant$ $leptogenesis$ mechanism \cite{Asaka:2018hyk,Iso:2010mv,Iso:2013lba,Granelli:2020ysj}.
\par
In this paper, we construct a modular model based on $A_4$ groups capable of explaining neutrino masses and mixing as well as the observed baryon asymmetry of the universe. For realization of inverse-seesaw, we extend the SM particle content with two left-handed neutral  fields $S_{i}$, two right-handed neutral fermion fields $N_{R_{i}}^c$ and a scalar field $\phi$. This gives us a specific structure of $7\times7$ neutrino mass matrix, which upon diagonalization gives seven mass eigenstates, three of them being the active neutrinos and the rest of them being heavy ones. Among the heavy neutrinos, we get two pairs of almost degenerate mass eigenstates and the decay of the lightest pair leads to resonant leptogenesis. We have also studied the charged lepton flavor violation (LFV) \cite{Davidson:2022jai} and neutrinoless double beta decay $(0\nu\beta\beta)$ \cite{Jones:2021cga} in our work.
\par
Our paper is organized as follows. In section \ref{sec2}, we explicitly addressed all the components of the model framework. Numerical analysis for the neutrino sector, neutrinoless double beta decay and charged lepton flavor violation is discussed in section \ref{sec3}. The study of resonant leptogenesis is carried out in section \ref{sec4} and finally we concluded our study in section \ref{sec5}.

%%%%%%%%%%%%%%%%%%%%% Model Framework %%%%%%%%%%%%%%%%%%%%%%%%%
%%%%%%%%%%%%%%%%%%%%%%%%%%%%%%%%%%%%%%%%%%%%%%%%%%%%%%%%%%%%%%%

\section{MODEL FRAMEWORK}\label{sec2}
In this section, we aimed to present the model framework, which is founded on $A_4$ modular symmetry. The charge assignment and modular weight of all the fields is shown in \textcolor{red}{Table I}. The particle content of the Standard Model is extended by two right-handed neutrino fields $N_{R_{i}}^c$ ($i = 1, 2$) and two singlets fermion $S_{i}$ ($i = 1, 2$) to implement inverse seesaw framework to study neutrino oscillation parameters and matter-antimatter asymmetry of our universe. We have taken the economical $(2, 2)$ inverse seesaw (ISS) setup to generate the neutrino masses in our work. To eliminate certain unwanted terms we adopt a local $U(1)_{B-L}$ symmetry which is spontaneously broken by VEV of scalar $\phi$. This spontaneous symmetry-breaking gives an massive gauge boson $Z'$ with mass $m_{Z'} = 2\sqrt{2}g_{\phi}v_{\phi}$. Here $g_{\phi}$ is gauge coupling and $\langle \phi \rangle = v_{\phi}$ is VEV of $\phi$ \cite{Iso:2010mv,Okada:2012fs}.

\begin{table}[h]
    \centering
    \begin{tabular}{||c|c|c|c|c|c|c|c|c|c|c|c|c||}
    \hline
    Symmetry & \multicolumn{12}{c||}{Superfield content and charge} \\
    \hline
    & $L_e$ & $L_\mu$ & $L_\tau$ & $e_R^c$ & $\mu_R^c$ & $\tau_R^c$ & $N_{R_1}^c$ & $N_{R_2}^c$ & $S_1$ & $S_2$ & $H_{u/d}$ & $\phi$ \\
    \hline
    $SU(2)_L$ & 2 & 2 & 2 & 1 & 1 & 1 & 1 & 1 & 1 & 1 & 1 & 1 \\
    $U(1)_Y$ & -1 & -1 & -1 & 2 & 2 & 2 & 0 & 0 & 0 & 0 & 1/-1 & 0 \\
    $U(1)_{B-L}$ & 0 & 0 & 0 & -1 & -1 & -1 & 1 & 1 & 0 & 0 & 1 & 1 \\
    $A_4$ & $1'$ & $1''$ & 1 & $1''$ & $1'$ & 1 & $1'$ & $1''$ & $1''$ & $1'$ & 1 & 1 \\
    $k_{I}$ & -1 & -1 & -1 & 1 & 1 & 1 & -3 & -3 & -2 & -2 & 0 & -3 \\
    \hline
    \end{tabular}
    \caption{Superfield content of the model with their $A_4$ and modular charges $k_I$.}
    \label{tab:particle-content}
\end{table}

\begin{table}[h]
    \centering
    \begin{tabular}{||c|c|c|c|c|c||}
    \hline
    & \multicolumn{5}{c||}{Yukawa Couplings} \\ \hline
    & $Y_{1}^{(4)}$ & $Y_{1'}^{(4)}$ & $Y_{1}^{(8)}$ & $Y_{1'}^{(8)}$ & $Y_{1''}^{(8)}$ \\ \hline
    $A_4$ & $1$ & $1'$ & $1$ & $1'$ & $1''$ \\ 
    $k_{I}$ & $4$ & $4$ & $8$ & $8$ & $8$ \\ \hline
    \end{tabular}
    \caption{Yukawa couplings with their $A_4$ and modular charges $k_I$.}
    \label{tab:yukawa-content}
\end{table}

We have given zero modular weights to the Higgs superfields $H_u$ and $H_d$. To make the charged lepton mass matrix diagonal we have assigned $-1$ modular to the lepton superfields $L_i$ $(i=e, \mu, \tau)$ and $+1$ modular weights to the chiral superfields corresponds to right-handed charged lepton $(e_R^c, \mu_R^c, \tau_R^c)$, which all are singlets under $A_4$ representations. The singlet representation of $(L_e, L_{\mu}, L_{\tau})$ are $(1', 1'', 1)$ and that of $(e_R^c, \mu_R^c, \tau_R^c)$ are $(1'', 1', 1)$. This ensures that the leptonic mixing matrix can originate entirely from the diagonalization of the neutrino mass matrix. To achieve $(2, 2)$ ISS setup we have incorporated two extra heavy right-handed fields with modular weights $-3$ each and they transform as $1'$ and $1''$ under $A_4$. Another pair of singlets superfield $S_{i}$ $(i=1,2)$ is also considered to obtain the Majorana mass term with modular weight $-2$ each and they transform under $A_4$ as $1''$ and $1'$ respectively. 

\par

The charge assignment is shown in \textcolor{red}{Table I} gives the following charged lepton superpotential:
\begin{equation}
    \mathcal{W}_l = y_{e} L_{e} H_{d} e_{R}^c + y_{\mu} L_{\mu} H_{d} \mu_{R}^c + y_{\tau} L_{\tau} H_{d} \tau_{R}^c
\end{equation}
The charged lepton mass matrix takes the diagonal form:
\begin{equation}
    M_{l} = \frac{v_d}{\sqrt{2}}\begin{pmatrix}
        y_{e} & 0 & 0\\
        0 & y_{\mu} & 0\\
        0 & 0 & y_{\tau}
    \end{pmatrix}
    =
    \begin{pmatrix}
        m_e & 0 & 0\\
        0 & m_{\mu} & 0\\
        0 & 0 & m_{\tau}
    \end{pmatrix}
\end{equation}

Here, $\langle H_{d} \rangle = \frac{v_d}{\sqrt{2}}$ is the VEV of Higgs superfield $H_{d}$. The Yukawa couplings can be adjusted to achieve the observed charged lepton masses $m_e$, $m_{\mu}$ and $m_{\tau}$.

\par

The invariant Dirac superpotential involving lepton fields $L_i$  and heavy right-handed fields $N_{R_{i}}^c$ is given as follows: 
\begin{equation}
    \mathcal{W}_{D} = \alpha_{p}[Y_{1'}^{(4)}L_{e}H_{u}N_{R_{1}}^c + Y_{1}^{(4)}L_{e}H_{u}N_{R_2}^c + Y_{1}^{(4)}L_{\mu}H_{u}N_{R_1}^c + Y_{1'}^{(4)}L_{\tau}H_{u}N_{R_2}^c]
\end{equation}
The Dirac mass matrix is obtained to be:
\begin{equation}
    M_{D} = \alpha_{p}\frac{v_u}{\sqrt{2}}\begin{pmatrix}
        Y_{1'}^{(4)} & Y_{1}^{(4)}\\
        Y_{1}^{(4)} & 0 \\
        0 & Y_{1'}^{(4)}
    \end{pmatrix}
\end{equation}
Here, $\alpha_{p}$ is free parameter and $\langle H_u \rangle = \frac{v_u}{\sqrt{2}}$ is the VEV of Higgs superfield $H_u$. The Yukawa couplings $Y_{1}^{(4)}$ and $Y_{1'}^{(4)}$ are transform under $A_4$ as $1$ and $1'$ respectively with modular weight 4 (See Appendix:\ref{sec: Append B}).

\par

The superpotential involving mixing between $N_{R_i}^c$ and sterile fermion field $S_i$ is given by:
\begin{equation}
    \mathcal{W}_{R} = \beta_{p}[Y_{1}^{(8)}N_{R_1}^cS_1 + Y_{1'}^{(8)}N_{R_1}^cS_2 + Y_{1''}^{(8)}N_{R_2}^cS_1 + Y_{1}^{(8)}N_{R_2}^cS_2]\phi
\end{equation}
This gives the following Majorana mass matrix:
\begin{equation}
    M_{R} = \beta_{p}v_{\phi}\begin{pmatrix}
        Y_{1}^{(8)} & Y_{1'}^{(8)}\\
        Y_{1''}^{(8)} & Y_{1}^{(8)}
    \end{pmatrix}
\end{equation}
Here, $\beta_{p}$ is the free parameter. The Yukawa couplings $Y_{1}^{(8)}$, $Y_{1'}^{(8)}$ and $Y_{1''}^{(8)}$ are transform under $A_4$ as $1$, $1'$ and $1''$ respectively with modular weight 8 (See Appendix:\ref{sec: Append B}).

\par

The invarient Majorana mass term for $S_i$ is given by the following superpotential:
\begin{equation}
    \mathcal{W}_{\mu} = \mu_{0}[Y_{1}^{(4)}S_1S_2 + Y_{1}^{(4)}S_2S_1 + Y_{1'}^{(4)}S_2S_2]
\end{equation}
The Majorana mass matrix is found to be:
\begin{equation}
    \mu = \mu_{0}\begin{pmatrix}
        0 & Y_{1}^{(4)}\\
        Y_{1}^{(4)} & Y_{1'}^{(4)}
    \end{pmatrix}
\end{equation}
Here, $\mu_{0}$ is free parameter.

\par

Within the current model, which incorporates \(A_4\) modular symmetry, the complete \(7 \times 7\) neutral fermion mass matrix for the inverse seesaw mechanism in the flavor basis \((\nu_{L}, N_{R}^c, S)\) is given by

\begin{eqnarray}
   \mathbb{M}=  \begin{pmatrix}
        0&M_D&0\\M_D^T&0&M_R^T\\0&M_R&\mu
    \end{pmatrix}.\label{eq:M matrix}
\end{eqnarray}

Using the appropriate mass hierarchy among the mass matrices as provided below:
\begin{equation}
    M_{R}>>M_{D}, \, \mu.
\end{equation}
The light neutrino mass in inverse seesaw is given by:
\begin{equation}
    m_{\nu} = M_D{M_R}^{-1}\mu{M_R^{-1}}^TM_D^T.
    \label{eq:nu mass}
\end{equation}
%%%%%%%%%% Numerical Analysis %%%%%%%%%%%%%%%%%%%%%%%%%%%%%%%

\section{Numerical analysis}\label{sec3}
\subsection{Neutrino fit data}

The neutrino mass matrix obtain in Eq.\eqref{eq:nu mass} is diagonalized by using the relation: $U^{\dagger}\mathcal{M}U = diag(m_{1}^{2}, m_{2}^{2}, m_{3}^{2})$, where $\mathcal{M} = m_{\nu}m_{\nu}^{\dagger}$ and U is a unitary matrix. Since the charged lepton mass matrix is diagonal in our study, $U=U_{PMNS}$ is the well-known PMNS matrix. This matrix is parametrized by three mixing angles $(\theta_{12}, \theta_{13}, \theta_{23})$, one Dirac phase ($\delta_{CP}$) and two Majorana phases ($\alpha_{21}, \alpha_{31}$) as given below.

\begin{equation}
 U_{PMNS} = \begin{pmatrix}
     c_{12}c_{13} & s_{12}c_{13} & s_{13}e^{-i\delta_{CP}}\\
     - s_{12}c_{23} - c_{12}s_{23}s_{13}e^{i\delta_{CP}} & c_{12}c_{23} - s_{12}s_{23}s_{13}e^{i\delta_{CP}} & s_{23}c_{13}\\
     s_{12}s_{23} - c_{12}c_{23}s_{13}e^{i\delta_{CP}} & - c_{12}s_{23} - s_{12}c_{23}s_{13}e^{i\delta_{CP}} & c_{23}c_{13}
 \end{pmatrix}  
 . \begin{pmatrix}
     1 & 0 & 0\\
     0 & e^{i\frac{\alpha_{21}}{2}} & 0\\
     0 & 0 & e^{i\frac{\alpha_{31}}{2}}
 \end{pmatrix}
\end{equation}
where, \(c_{ij}=\cos{\theta_{ij}}\) and \(s_{ij}=\sin{\theta_{ij}}\). Using the PMNS matrix, we can calculate the three neutrino mixing angles as follows:
\begin{equation}
    \sin^2{\theta_{13}}=|U_{13}|^2, \, \sin^2{\theta_{12}} = \frac{|U_{12}|^2}{1 - |U_{13}|^2}, \, \sin^2{\theta_{23}}=\frac{|U_{23}|^2}{1 - |U_{13}|^2}.
\end{equation}
The Jarlskog invariant $J_{CP}$ and the CP-violating phase $\delta_{CP}$ are calculated using the elements of the PMNS matrix as given below:
\begin{equation}
    J_{CP} = Im[U_{e1}U_{\mu2}U_{e2}^{*}U_{\mu1}^{*}] = s_{23}c_{23}s_{12}c_{12}s_{13}c_{13}^{2}\sin{\delta_{CP}}.
\end{equation}
The Majorana phases are also calculated using the following two invariants:
\begin{equation}
\begin{split}
& I_{1} = Im[U_{e1}^{*}U_{e2}] = c_{12}s_{12}c_{13}^{2}\sin{\left(\frac{\alpha_{21}}{2}\right)},\\
& I_{2} = Im[U_{e1}^{*}U_{e3}] = c_{12}s_{13}c_{13}\sin{\left(\frac{\alpha_{31}}{2}-\delta_{CP}\right)}.
\end{split}    
\end{equation}

In numerical analysis, we use the global fit neutrino oscillation data within a $3\sigma$ interval as given in Table \ref{tab:oscillation data}.
\begin{table}[h]
    \centering
    \begin{tabular}{||c||c|c||}
    \hline
    \multicolumn{3}{|c||}{Normal ordering}\\
    \hline
           & bfp $\pm 1\sigma$ & $ 3\sigma$ range \\
         \hline
         $\sin^2{\theta_{12}}$ & $0.303_{-0.011}^{+0.012}$  & $0.270 - 0.341$ \\
         $\sin^2{\theta_{13}}$& $0.02203_{-0.00059}^{+0.00056}$ & $0.02029 - 0.02391$ \\
         $\sin^2{\theta_{23}}$& $0.572_{-0.023}^{+0.018}$ & $0.406 - 0.620$  \\
         $\frac{\Delta m_{21}^2}{10^{-5} (eV^2)}$ & $7.41_{-0.20}^{+0.21}$ & $6.82 - 8.03$  \\
         $\frac{\Delta m_{31}^2}{10^-{3} (eV^2)}$& $2.511_{-0.027}^{+0.028}$ & $2.428 - 2.597$  \\
    \hline
    \end{tabular}
    \caption{The NuFIT 5.2 (2022) results \cite{Esteban:2020cvm}.}
    \label{tab:oscillation data}
\end{table}

\begin{figure}[!ht]
        \centering
        \includegraphics[scale=0.45]{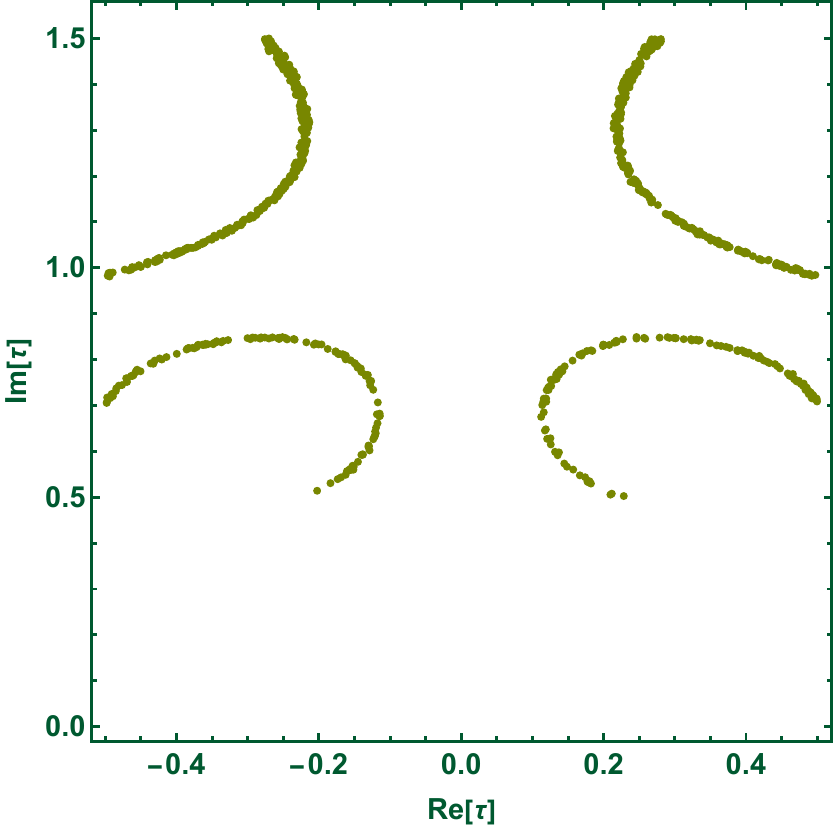}
        \hspace{2em}
        \includegraphics[scale=0.45]{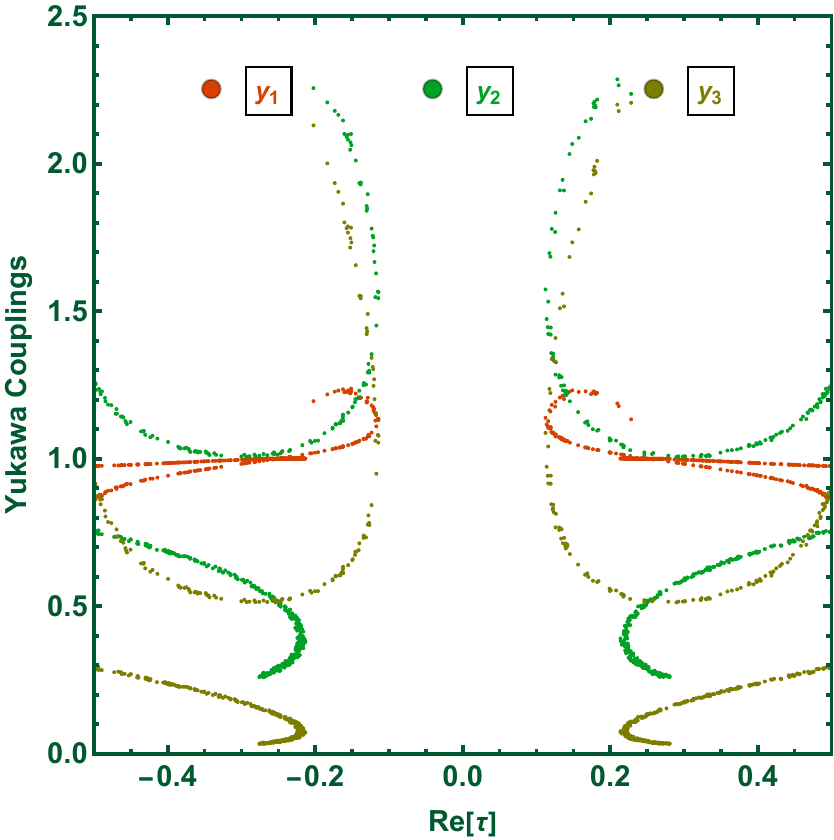}
        \includegraphics[scale=0.45]{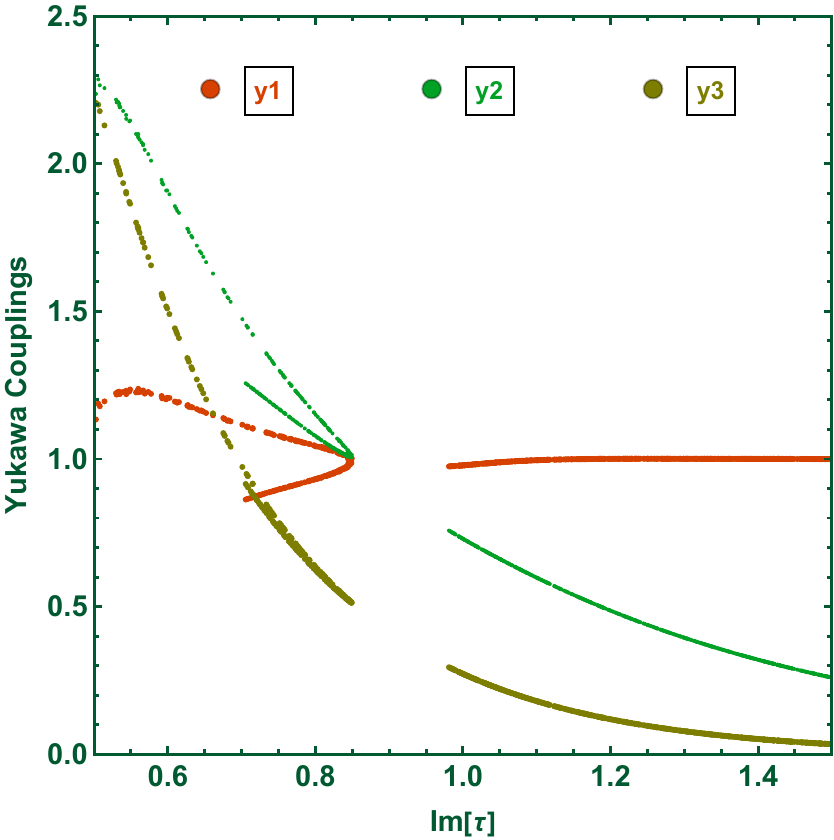}
        \caption{($Top-Left$) Correlation between the real component and the imaginary component of $\tau$. ($Top-Right$) Variation of Yukawa couplings ($y_{1}, y_{2}, y_{3}$) with Re[$\tau$]. ($Bottom$) The variation of Yukawa couplings ($y_{1}, y_{2}, y_{3}$) with Im[$\tau$].}
       \label{fig:fig-Yukawa and Tau} 
\end{figure}
\begin{figure}
    \centering
    \includegraphics[scale=0.45]{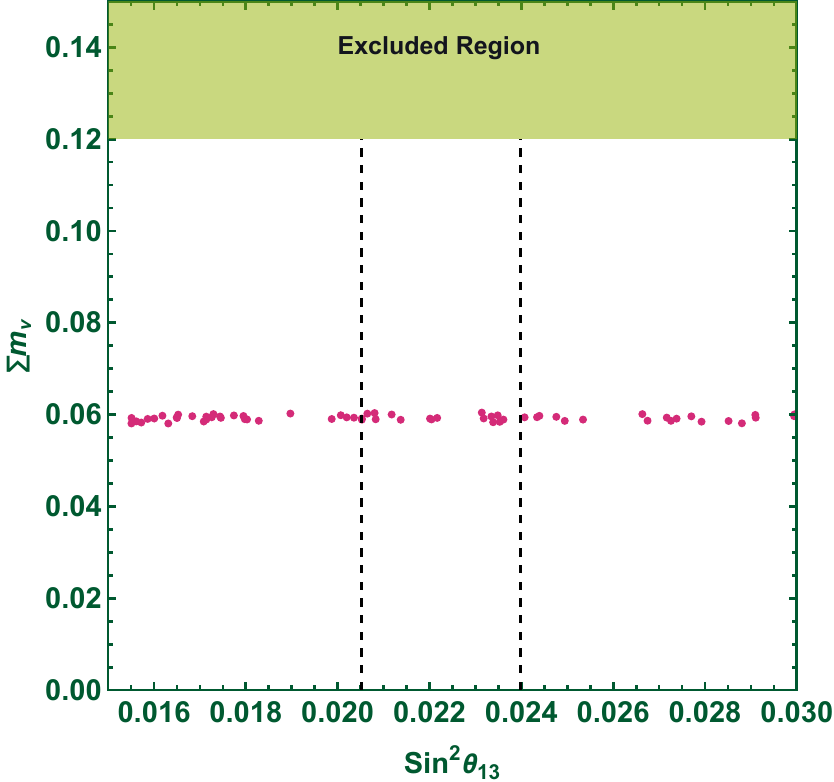}
    \hspace{2em}
    \includegraphics[scale=0.45]{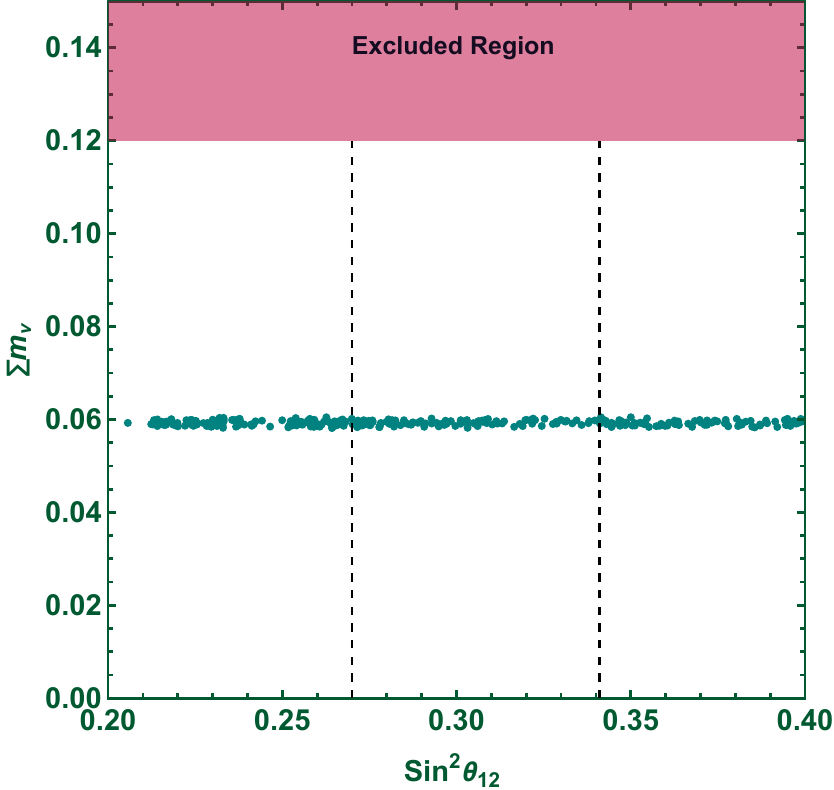}
    \includegraphics[scale=0.45]{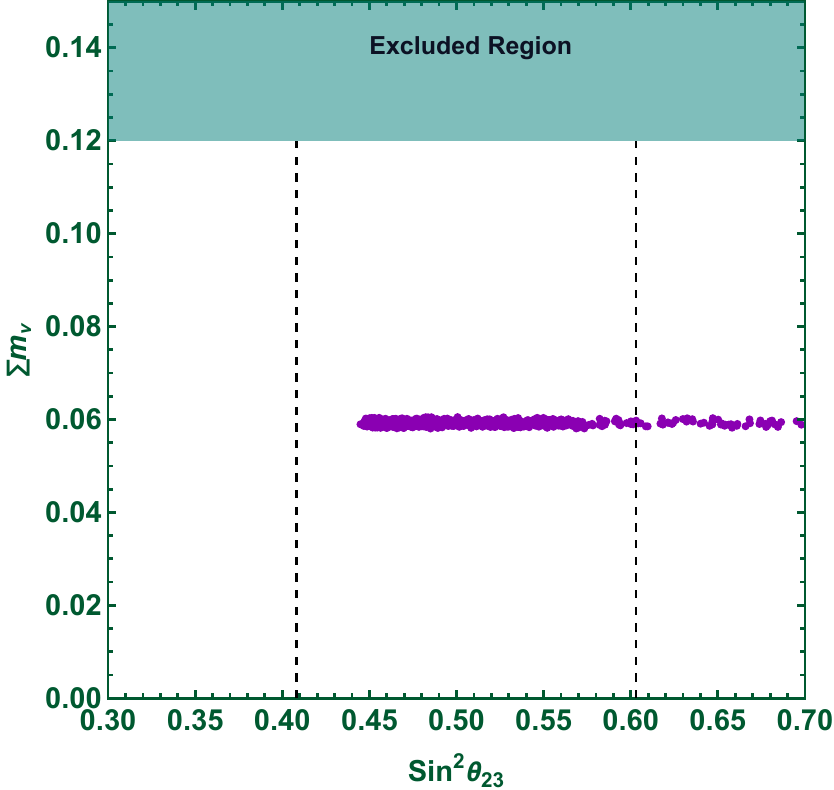}
    \caption{Variation of sum of the neutrino masses $\sum m_{\nu}$ in eV with the mixing angles $\sin^{2}{\theta_{13}}, \sin^{2}{\theta_{12}}$ and $\sin^{2}{\theta_{23}}$. }
    \label{fig:sum nu vs angle}
\end{figure}
To match the current neutrino oscillation data, we select the following ranges for the model parameters:
\begin{equation}
\begin{split}
& \text{Re}[\tau] \in [-0.5, 0.5], \quad \text{Im}[\tau] \in [0.5,1.5], \quad\alpha_{p} \in [10^{-4},10^{-3}],\\
& \beta_{p} \in [10^{-3},10^{-2}], \quad \mu_{0} \in [0.1, 10] \, GeV, \quad v_{\phi} \in [10^5, 10^8]\, GeV.
\end{split}
\end{equation}
The vacuum expectation values (VEVs) of the Higgs superfields, \( H_u \) and \( H_d \), are given by \( \langle H_u \rangle = \frac{v_u}{\sqrt{2}} \) and \(\langle H_d \rangle = \frac{v_d}{\sqrt{2}} \). These are related to the Standard Model Higgs VEV, \( v_H \), through the relation \( v_H = \sqrt{v_u^2 + v_d^2} \) and the ratio of their VEVs is expressed as \( \tan\beta = \frac{v_u}{v_d} = 5 \) \cite{Kashav:2021zir,Antusch:2013jca}. 
The input parameters are randomly scanned within the specified parameter ranges. The permitted regions are filtered based on the \(3\sigma\) limits of solar and atmospheric mass squared differences and mixing angles provided in the Table.\ref{tab:oscillation data}. The correlation between \(\text{Re}[\tau]\) and \(\text{Im}[\tau]\) is shown in the Fig.\ref{fig:fig-Yukawa and Tau}. The Fig.\ref {fig:fig-Yukawa and Tau} also shows the correlation between the components of the Yukawa couplings $y_{1}$, $y_2$, and $y_3$ \footnote{$y_1$, $y_2$ and $y_3$ are components of Yukawa coupling Y which transform as a 3 under $A_4$.(See Appendix:\ref{sec: Append B}).} with the Real and Imaginary part of modulus $\tau$.

\begin{figure}
    \centering
    \includegraphics[scale=0.45]{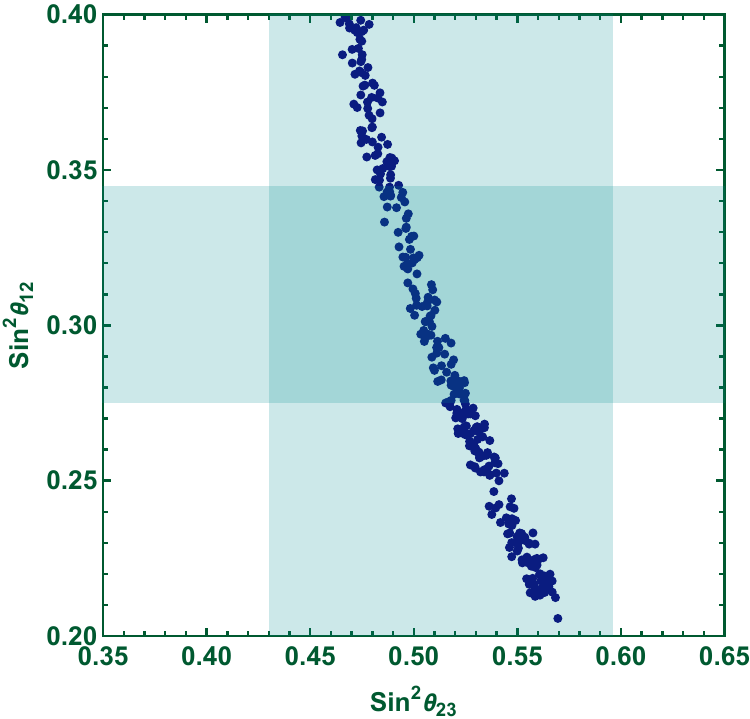}
    \caption{Correlation between mixing angle $\sin^{2}{\theta_{23}}$ and $\sin^{2}{\theta_{12}}$}
    \label{fig:s23 and s13 relation}
\end{figure}

Fig.\ref{fig:sum nu vs angle} illustrates variation in the sum of neutrino masses with the neutrino mixing angles. The sum of the neutrino masses is found to be below the upper bound $\sum m_{\nu} < 0.12$\, eV provided by Planck 2018\cite{Planck:2018vyg}. From the bottom panel of Fig.\ref{fig:sum nu vs angle}, it can be observed that our model predicts a lower limit of 0.44 on the values of atmospheric mixing angle $\sin^{2}{\theta_{23}}$ within $3\sigma$ range. Our model predict strong correlation between the mixing angle $\sin^{2}{\theta_{23}}$ and $\sin^{2}{\theta_{12}}$. It is evident from Fig.\ref{fig:s23 and s13 relation}. Our model suggests a linear relationship between these two mixing angles in the $3\sigma$ region.

%%%%%%%%%%%%%%%%%%%%%%%%%%%%%%%%%%%%%%%%%%%%%%%%%%%%%%%%%%
\begin{figure}[!ht]
    \centering
        \centering
        \includegraphics[scale=0.45]{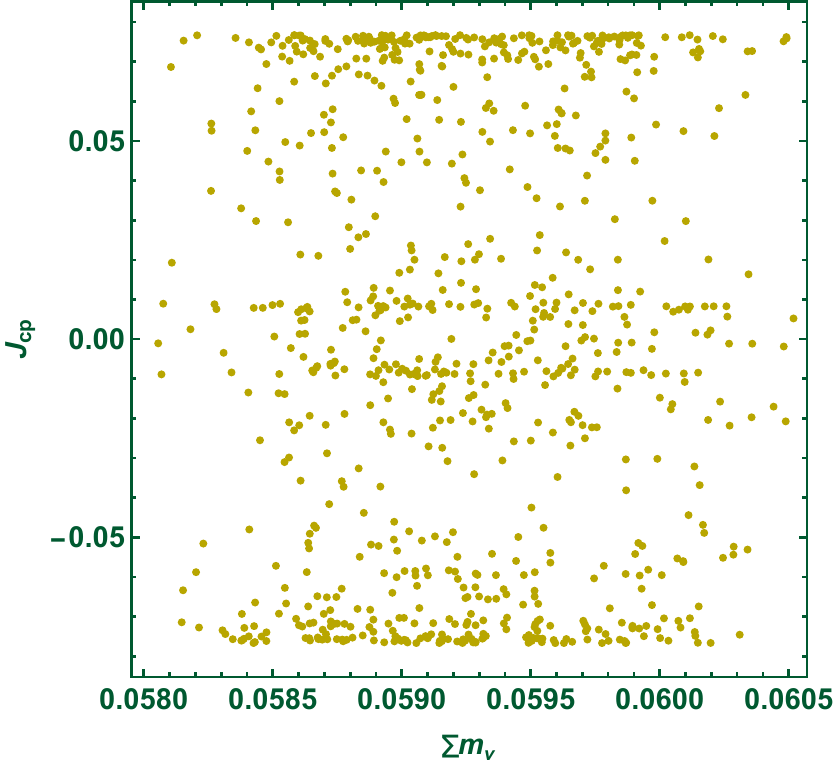}
        \includegraphics[scale=0.48]{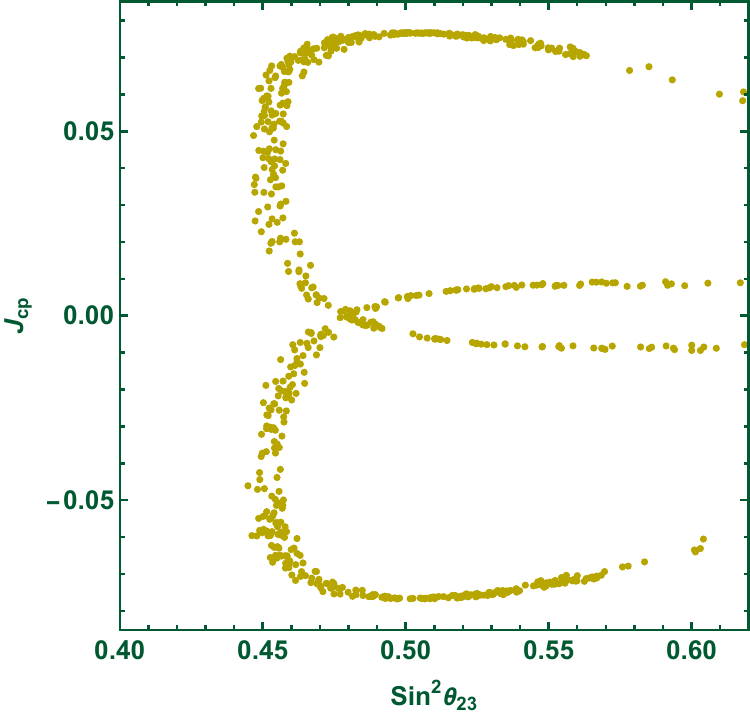}
        \caption{($Left$) Correlation between $J_{CP}$ and sum of neutrino masses $(\sum m_{\nu})$. ($Right$) Variation of $J_{CP}$ with the mixing angle $\sin^{2}{\theta_{23}}$.}
        \label{fig:fig-jcp}
\end{figure}
%%%%%%%%%%%%%%%%%%%%%%%%%%%%%%%%%%%%%%%%%%%%%%%%%%%%%%%%%%%%%%%

\begin{figure}[!ht]
        \centering
        \includegraphics[scale=0.45]{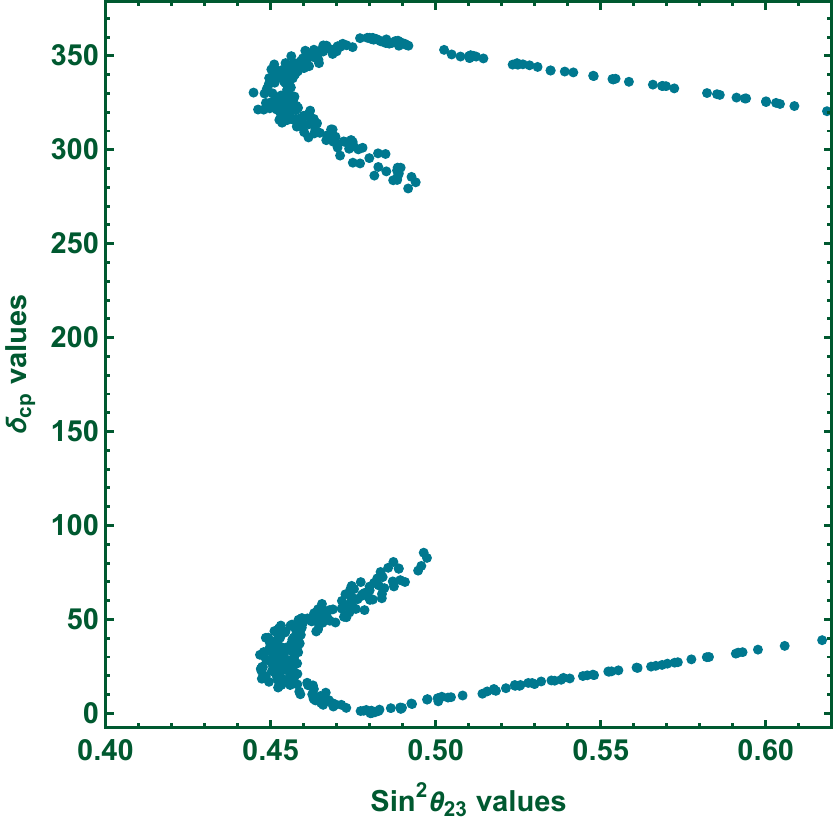}
        \includegraphics[scale=0.45]{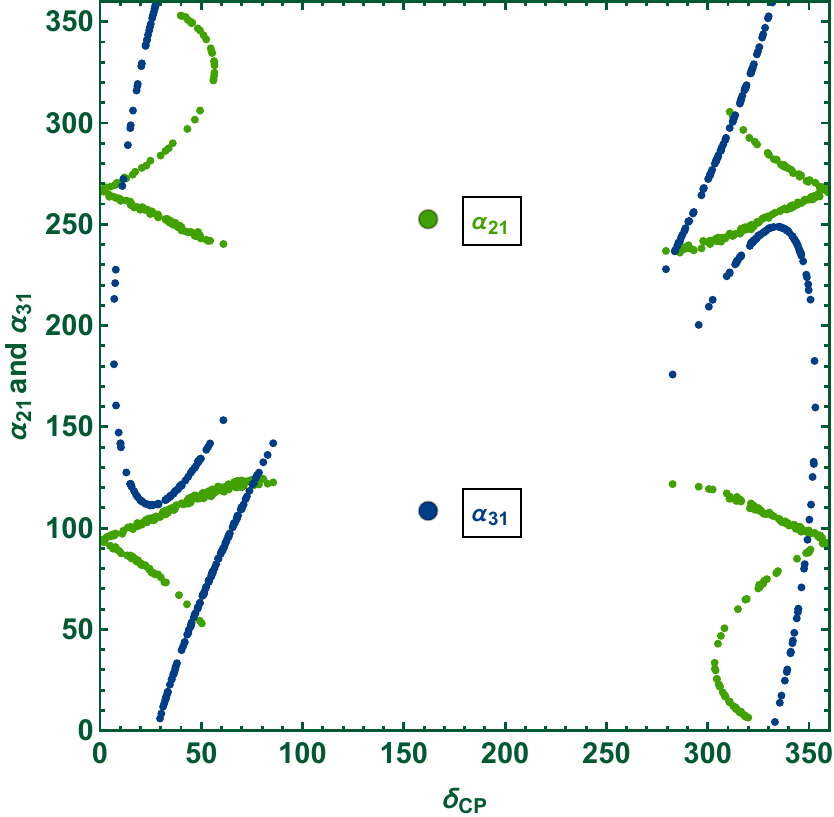}
        \caption{($Left$) Correlation between the Direc CP phase with the Sine square of the atmospheric mixing angle $(\sin^{2}{\theta_{23}})$. ($Right$) Interdependence of Majorna phases (\(\alpha_{21}\) \& \(\alpha_{31}\)) with the Dirac CP phase $(\delta_{CP})$.}
        \label{fig:fig-majoran phase}
\end{figure}

In Fig.\ref{fig:fig-jcp}, we have shown the relation of Jarlskog invariant $J_{CP}$ with the sum of neutrino mass ($left-panel$) and with $\sin^{2}{\theta_{23}}$ ($right-panel$). The value of $J_{CP}$ is found to be in the range $-0.07\lesssim J_{CP}\lesssim 0.07$. We have also shown the interdependence between $\delta_{CP}$ and $\sin^{2}{\theta_{23}}$ in the $left-panel$ of Fig.\ref{fig:fig-majoran phase}. The value of Delta CP phase $\delta_{CP}$ is found in the range $\delta_{CP}^{o}\in[0^{\circ},85^{\circ}]$ and $\delta_{CP}^{o}\in[279^{\circ},359^{\circ}]$. In the $right-panel$ of Fig.\ref{fig:fig-majoran phase}, the correlation between the two Majorana phases ($\alpha_{21}$, $\alpha_{31}$) and $\delta_{CP}$ is shown.

%%%%%%%%%%%%%%%%%%%   Detections    %%%%%%%%%%%%%%%%%%%%%%%%%%

%%%%%%%%%%%%%% Neutrinoless double beta decay %%%%%%%%%%%%%%%%%%%%%

\subsection{Neutrinoless Double Beta Decay ($0 \nu \beta \beta$)}
\begin{figure}[!ht]
    \centering
        \centering
        \includegraphics[scale=0.5]{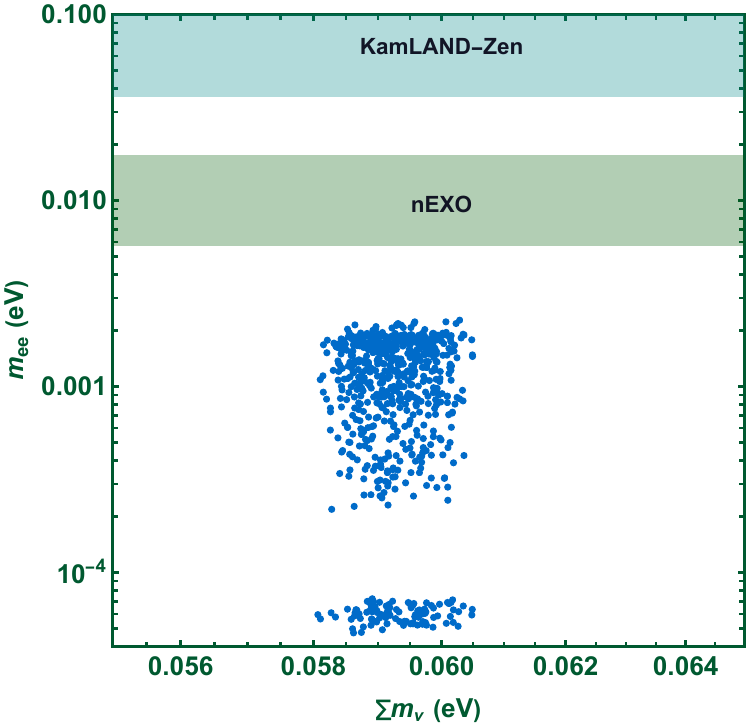}
        \caption{Effective neutrino mass $|m_{ee}|$ as a function of sum of neutrino mass ($\sum m_{\nu}$). The shaded region represents the current experimental limit on effective neutrino mass.}
        \label{fig:fig-NDBD}
\end{figure}

Neutrinoless double beta decay is a hypothetical nuclear reaction in which a nucleus undergoes double beta decay without emitting neutrinos:
\begin{equation}
    (Z, A) \longrightarrow (Z+2,A) + 2e^{-}
\end{equation}
These reactions violate conservation of lepton number by $\Delta L=2$ units. Its detection would confirm that neutrinos are Majorana particles. The decay rate is proportional to the effective neutrino mass $|m_{ee}|$:
\begin{equation}
    |m_{ee}|=|m_{1}\cos^{2}{\theta_{12}}\cos^{2}{\theta_{13}} + m_{2}\sin^{2}{\theta_{12}}\cos^{2}{\theta_{13}}e^{i \alpha_{21}} + m_{3}\sin^{2}{\theta_{13}}e^{i(\alpha_{31}-2\delta_{CP})}|
   \label{NDBD} 
\end{equation}

The current experimental limit from KamLAND-Zen \cite{KamLAND-Zen:2022tow} experiment is $(36-156)$ meV. The sensitivity limits on $|m_{ee}|$ projected for future experiment nEXO \cite{nEXO:2017nam} lie within $(5.7-17.7)$ meV at $90\%$ C.L. We have illustrated $|m_{ee}|$ as a function of sum of the neutrino mass $\sum m_{\nu}$ in Fig.\ref{fig:fig-NDBD}.
%%%%%%%%%%%%%%%%%%%%%%%%%%%%%%%%%%%%%%%%%%%%%%%%%%%%%%%%%%%%%%%%
%%%%%% Lepton Flavour Violation %%%%%%%%%%%%%%%%%%%%%%%%%%%%

\subsection{Lepton Flavor Violation}
There is clear evidence of lepton flavor violation in the neutrino sector, but no such violation has been observed in the charged lepton sector so far. If observed, this would be a clear signal of physics beyond the Standard Model (BSM).  Lepton flavor-violating decays such as \(\mu \rightarrow e + \gamma\), \(\mu \rightarrow eee\), and \(\mu \rightarrow e\) conversion in nuclei, which are suppressed in the Standard Model by the GIM mechanism, could have significant contributions in the current model \cite{Calibbi:2017uvl, Ardu:2022sbt}.


\begin{figure}[!ht]
    \centering
     \includegraphics[width=0.45\textwidth]{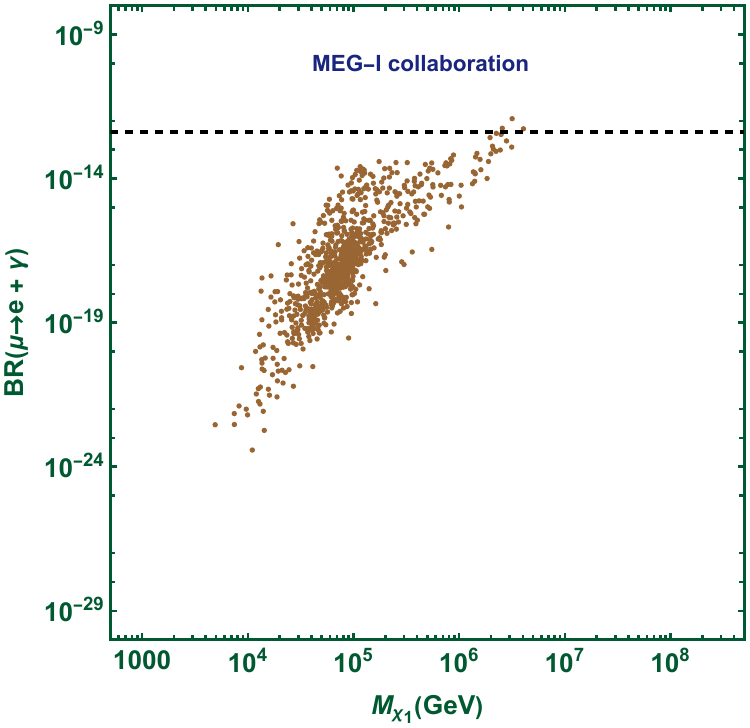}
     \hspace{2em}
     \includegraphics[width=0.45\textwidth]{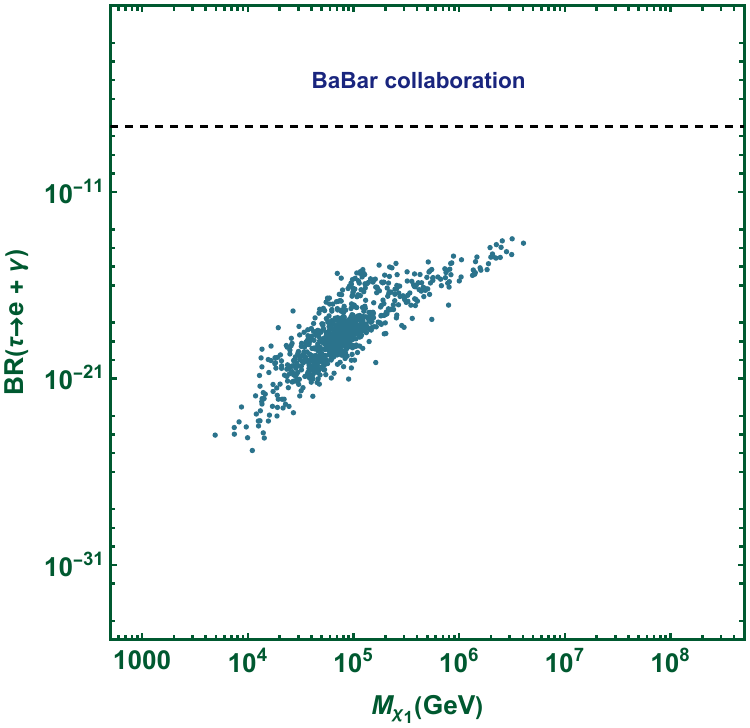}
     \includegraphics[width=0.45\textwidth]{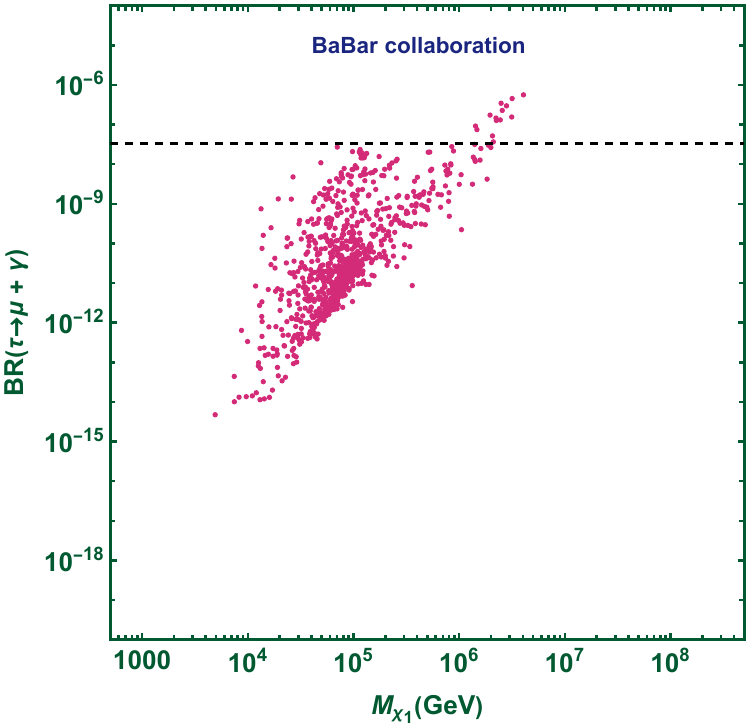}
        \caption{Relationship between $M_{{\chi}_{1}}$ and branching ratio (BR) of different LFV processes.}
        \label{fig:fig-13}
\end{figure}
The $7 \times 7 $ matrix \(\mathbb{M}\) in Eq. \eqref{eq:M matrix} can be diagonalised as follows:
\begin{equation}
    U^{T}\mathbb{M}U = diag(m_{1}, m_{2}, m_{3}, M_{{\chi}_{1}}, M_{{\chi}_{2}}, M_{{\chi}_{3}}, M_{{\chi}_{4}})
\label{eq:LFV data}    
\end{equation}
where $U$ is the $7 \times 7$ unitary matrix. Here, $m_{1}, m_{2}, m_{3}$ are the mass of active neutrinos and $M_{{\chi}_{i}}$'s ($i=1,2,3,4$) are the mass of the heavy neutrino corresponding to four heavy mass eigen states $\chi_{i}$. 
\par
The lepton flavor-violating decays $l_{i}\rightarrow l_{j}\gamma$ can occur through the exchange of heavy fermions at the one-loop level. This is due to the mixing between light and heavy fermions. The leading one-loop contribution to the branching ratios for these decays is described by the dominant term, which is given as follows\cite{Chakraborty:2021azg, Deppisch:2004fa, Forero:2011pc, Chekkal:2017eka, Ilakovac:1994kj}:
\begin{equation}
    BR(l_{i}\longrightarrow l_{j}\gamma)=\frac{\alpha^{3}\sin^2{\theta_{W}}}{256\pi^{2}}\left(\frac{M_{l_{i}}}{M_{W}}\right)^{4}\frac{M_{l_{i}}}{\Gamma_{l_{i}}}|F_{ij}|^{2}
\end{equation}
where \(F_{ij}\) is loop function whose analytical form is:
\begin{equation}
 \begin{aligned} F_{ij}=\sum_{k}U_{ik}^{*}U_{jk}F_{\gamma}\left(\frac{M_{{\chi}_{i}}^{2}}{M_{W}^{2}}\right),\text{with}\\
 F_{\gamma}(x)=-\frac{2x^{3}+5x^{2}-x}{4(1-x)^{2}}-\frac{3x^{3}}{2(1-x)^{4}}\ln{x}.
 \end{aligned}      
\end{equation}
where \(\alpha\) is the fine structure constant, \(\theta_{W}\) is the Weinberg angle, $\Gamma_{i}$ is the decay width of $i$th lepton, \(U_{ij}\) is the $ij^{th}$ element of the unitary matrix $U$, \(M_{W}\) is the mass of $W$ boson, \(M_{{\chi}_{i}}\)'s are the masses of heavy fermions and \(M_{l_{i}}\) is the mass of decaying lepton.  The decay width of $\mu$ and \(\tau\) are $3.01 \times 10^{-19}$ GeV and \(2.267\times 10^{-12}\) GeV respectively \cite{Ilakovac:1994kj,Chakraborty:2021azg}.
\par
Fig. \ref{fig:fig-13} shows the branching ratio for the three processes as denoted by the key text. The horizontal line in $top-left$ panel stands for MEG-I collaboration bound \(4.2 \times 10^{-13}\) \cite{MEG:2016leq,Baldini:2013ke} for the BR$(\mu\rightarrow e\gamma)$. The horizontal line in $top-right$ panel and $bottom$ panel represents the experimental upper bound \(3.3 \times 10^{-8}\) set by BaBar collaboration\cite{BaBar:2009hkt} for the $\tau$-decay processes. 

%%%%%%%%%%%%%----------BAU--------------------------------%%%%%%%%%%%

\section{Resonant leptogenesis}
\label{sec4}
In our work, we study baryogenesis via the mechanism known as leptogenesis. This mechanism is fascinating because it establishes a link between the properties of neutrinos and the origin of the baryon asymmetry. The lepton asymmetry of the universe is produced by CP violating out-of-equilibrium decays of heavy right-handed neutrinos, the seesaw companion of ordinary neutrinos. The relation between the baryon asymmetry and a primordial $(B - L)$ asymmetry is given as follows\cite{Khlebnikov:1988sr}:
\begin{equation}
    Y_B = \left(\frac{8N_f + 4N_H}{22N_f + 13N_H}\right) Y_{B-L},
    \label{Eq of YB}
\end{equation}
where $N_H$ is the number of Higgs doublets and $N_f$ is the number of families of fermion.\\
Taking into account the effect of the two Higgs doublets, the baryon asymmetry can be calculated in our study as $Y_{B}=8/23Y_{B-L}$\footnote{Here, $N_{f} = 3$, $N_{H} = 2$ and it is assumed that both Higgs doublets survive at sphaleron freeze-out temperature. In case of one Higgs double: $Y_{B} = 28/79Y_{B-L}$.}.
The required $(B-L)$ asymmetry can be generated as the lepton asymmetry as suggested by Fukugita and Yanagida \cite{Fukugita:1986hr}. The baryon asymmetry yield is then calculated by solving the Boltzmann equation. It is sufficient to consider the Boltzmann equation for the lightest right-handed neutrino when the mass of the three right-handed neutrinos is non-degenerate. Because the inverse decay process of the lightest right-handed neutrinos, before their out-of-equilibrium decay, washed out the lepton asymmetry produced by heavier right-handed neutrinos. However, in resonant leptogenesis, we consider the lightest pair of right-handed neutrinos to produce the desired CP violation, which will be discussed in the following section.
\par
As mentioned in the paper\cite{Marciano:2024nwm}, We can avoid the superparticles contribution to the final baryon asymmetry and the complication related to the gravitino densities assuming the mass scale of right-handed neutrino above the SUSY breaking scale $m_{SUSY}$. The correction to the lepton masses of the order $m_{SUSY}/\mathcal{F}$ is needed, where $\mathcal{F}$ is the scale at which the supersymmetry breaking sector communicates with the SM sector \cite{Criado:2018thu}. In our work, the order of right-handed neutrinos is $\mathcal{O}(10^5, 10^{7})$ GeV. Hence, if we take $m_{SUSY} =\mathcal{O} (10^{14})$GeV and $\mathcal{F} =\mathcal{O} (10^{18})$ GeV, then correction to the lepton masses do not affect the low energy results and we can carry out leptogenesis without considering any massive superparticles.
\par
The lepton asymmetry generated from the decay of right-handed neutrinos is diluted by scattering and decay processes, collectively known as \textit{washout processes}. The strength of washout is given by the following $K$ parameter
\begin{equation}
    K_{i}=\frac{\Gamma_{i}}{H(T=M_{\chi_i})}=\frac{m^{e}_{i}}{m^*}
\end{equation}
where $\Gamma_{i}$ is the decay rate of right-handed neutrinos. The effective neutrino mass is given by the following relations\cite{Buchmuller:1996pa,Davidson:2008bu}:
\begin{equation}
    m^{e}_{i}=\frac{(hh^{\dagger})_{ii}v_{u}^2}{M_{\chi_i}} 
\end{equation}
and $m^* \simeq 1.1*10^{-3}$\quad eV. The value of $K<<1$, $K \simeq 1$, and $K>>1$ refers to weak, intermediate, and strong washout regimes, respectively. The contribution of $\Delta L=1$ scattering processes between $\chi_{i}$ and leptons mediated by the Higgs boson $H_{u}$, involving top quarks or electroweak (EW) gauge bosons, and $\Delta L=2$ processes between the leptons and $H_{u}$ mediated by $\chi_{i}$ can be safely neglected in our study as in our work $K>>1$ \cite{Buchmuller:2003gz,Buchmuller:2002rq}. Therefore, we will only consider decay and $Z'$ mediated interactions in our BAU study.
\subsection{CP asymmetry}
\begin{figure}
        \centering
        \includegraphics[scale=0.45]{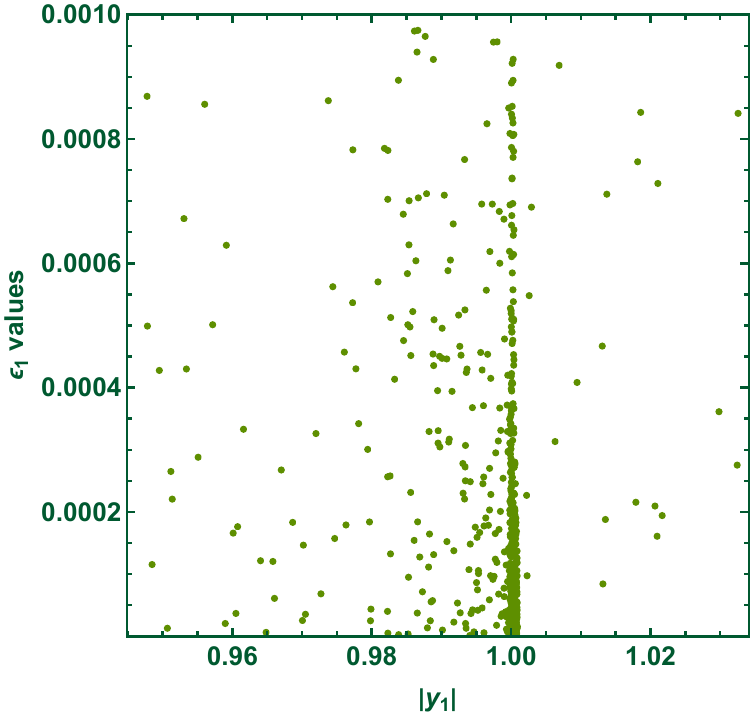}
        \hspace{2em}
        \includegraphics[scale=0.45]{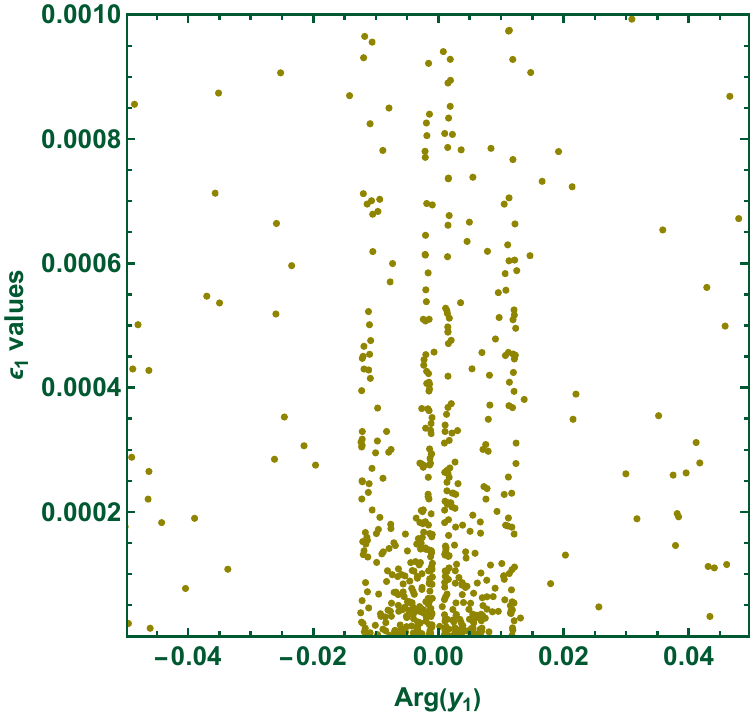}
        \includegraphics[scale=0.45]{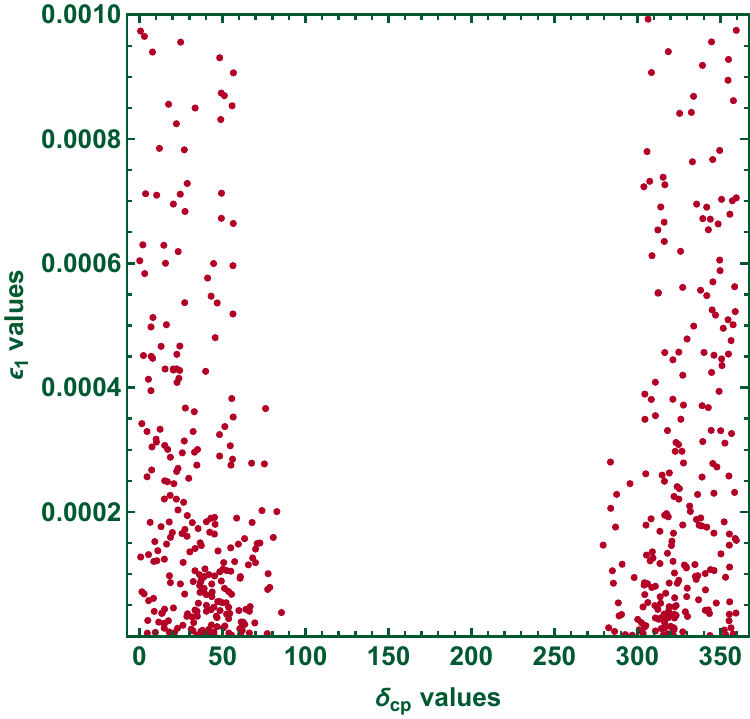}
        \caption{($Top-Left$) Correlation between absolute value of $y_{1}$ and CP asymmetry $\epsilon_{1}$. ($Top-right$) Relationship between Arg($y_{1}$) and CP asymmetry $\epsilon_{1}$. ($Bottom$) Variation of Dirac CP phase $\delta_{CP}$ with CP asymmetry $\epsilon_{1}$.}
       \label{fig:fig-CP} 
\end{figure}

In this section, we will calculate CP asymmetry produced from the decay of two degenerate right-handed neutrinos $\chi_1(\chi_2)$ into $\chi_2(\chi_1)$. Now, the lower block of matrix $\mathbb{M}$ in \eqref{eq:M matrix} can be written as $4\times4$ matrix:
\begin{equation}
    \mathbb{M}_H =\begin{pmatrix}
        0 & M_{R}^T\\
        M_R & \mu
    \end{pmatrix}
\end{equation}
After diagonalising $\mathbb{M}_H$ by $4\times4$ unitary matrix $W$, we get two degenerate pairs of mass namely $(\chi_1, \chi_2)$ and $(\chi_3, \chi_4)$.
\begin{equation}
    (\mathbb{M}_{H})_{diag} = W^{T}\mathbb{M}_{H}W = diag(\chi_1, \chi_2, \chi_3, \chi_4)
\end{equation}

\begin{figure}
        \centering
        \includegraphics[scale=0.65]{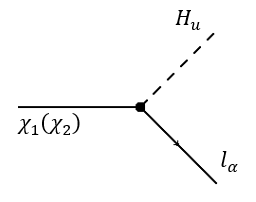}
        \includegraphics[scale=0.65]{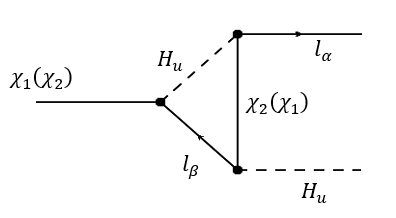}
        \includegraphics[scale=0.65]{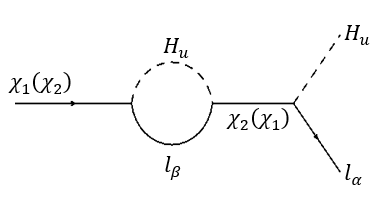}
        \caption{Diagram contributing to CP asymmetry $\epsilon_{1}$ and $\epsilon_{2}$.}
       \label{fig:fig-CP asymmetry} 
\end{figure}

For decay of $\chi_i$ into $l_{\alpha}H_u(\bar{l_{\alpha}}H^{\dagger}_u)$, the CP asymmetry $\epsilon_i$ is given by \cite{Iso:2013lba,Chakraborty:2021azg}:
\begin{eqnarray}
    \epsilon_i &=& \frac{\sum_{\alpha}[\Gamma(\chi_{i} \longrightarrow l_{\alpha}H_u) - \Gamma(\chi_i \longrightarrow\bar{l_{\alpha}}H^{\dagger}_u)]}{\sum_{\alpha}[\Gamma(\chi_i \longrightarrow l_{\alpha}H_u) + \Gamma(\chi_i \longrightarrow\bar{l_{\alpha}}H^{\dagger}_u)}\\
    &=&\frac{1}{8\pi}\sum_{i\neq j}\frac{Im[(hh^{\dagger})_{ij}^{2}]}{(hh^{\dagger})_{ii}}f_{ij}
\end{eqnarray}
where $\alpha=e, \mu, \tau$. Here, $f_{ij}$ represents the contribution to CP asymmetry from vertex correction and self-energy correction. In the case of resonant leptogenesis, $f_{ij}$ only gets a contribution from self-correction. 
\begin{equation}
    f_{ij}\approx f_{ij}^{self} = \frac{(M_{\chi_i}^2 - M_{\chi_j}^2)M_{\chi_i}M_{\chi_j}}{(M_{\chi_i}^2 - M_{\chi_j}^2)^2 + R_{ij}^2},   
\end{equation}
Where, the regulator $R_{ij}$ is given by following relation \cite{Iso:2013lba}:

\begin{equation}
    R_{ij} = |M_{\chi_i}\Gamma_i + M_{\chi_j}\Gamma_j|
\end{equation}
Here, Decay width of $\chi_i$ is given by:
\begin{equation}
    \Gamma_i = \frac{(hh^{\dagger})_{ii}M_{\chi_i}}{8\pi}
\end{equation}

In all of the above relations Yukawa couplings in diagonal mass basis $h_{i\alpha}$ are related to Yukawa couplings in flavor basis $y_{i\alpha}$ via the following relations \cite{Chakraborty:2021azg}:

\begin{eqnarray}
    h_{1\alpha}&=&V_{11}^{*}y_{1\alpha} + V_{12}^{*}y_{2\alpha}\\
    h_{2\alpha}&=&V_{21}^{*}y_{1\alpha} + V_{22}^{*}y_{2\alpha}\\
    h_{3\alpha}&=&V_{13}^{*}y_{1\alpha} + V_{23}^{*}y_{2\alpha}\\
    h_{4\alpha}&=&V_{14}^{*}y_{1\alpha} + V_{24}^{*}y_{2\alpha}
\end{eqnarray}

Since only the lightest pair $(\chi_1, \chi_2)$ decay will contribute to CP asymmetry generation, we can compute $\epsilon_1, \epsilon_2$ via the following relation \cite{Chakraborty:2021azg}:
\begin{eqnarray}
    \epsilon_1&=&\frac{1}{8\pi}Im[(hh^{\dagger})_{12}^{2}f_{12} + (hh^{\dagger})_{13}^{2}f_{13} + (hh^{\dagger})_{14}^{2}f_{14}]\\
    \epsilon_2&=&\frac{1}{8\pi}Im[(hh^{\dagger})_{21}^{2}f_{21} + (hh^{\dagger})_{23}^{2}f_{23} + (hh^{\dagger})_{24}^{2}f_{24}]
\end{eqnarray}

It is important to note that the CP asymmetry parameter depends on the Yukawa couplings apart from the VEV $v_{\phi}$ and free parameters. Since these Yukawa couplings are determined by the modulus \(\tau\) constrained by neutrino oscillation data, we have limited control over their magnitude. In our present model, CP asymmetry is generated by the vacuum expectation value of the modulus $\tau$. Also, this vacuum expectation value is related to the CP phase of the PMNS matrix in the neutrino sector. Therefore, we expect some correlation among Yukawa couplings, the Dirac CP phase, and the CP asymmetry of leptogenesis in our study. In Fig.\ref{fig:fig-CP}, we have shown the correlation between CP asymmetry and the absolute value of Yukawa couplings $y_{1}$\footnote{The Yukawa coupling of modular weight 2 is ($y_1,y_2,y_3$) and all other higher weight modular weight can be written in terms of these three Yukawas. In our study, we have used the modular form of weight 4 and 8, which can also be written in terms of ($y_1,y_2,y_3$) (See Appendix:\ref{sec: Append B}).} and its argument. Also, the relationship between the Dirac phase $\delta_{CP}$ and CP asymmetry $\epsilon_{1}$ is shown. It should be noted that in our study $\epsilon_{1} \approx \epsilon_{2}$.
%%%%%%%%----------Boltzmann Equations--------------%%%%%%%%%

\subsection{Boltzmann Equations}

If it is assumed that all the standard model particles are in thermal equilibrium, the Boltzmann equation for $\chi_1$, $\chi_2$ and the $(B-L)$ asymmetry can be written as \cite{Iso:2010mv}:

\begin{eqnarray}
    \frac{dY_{\chi_1}}{dz}&=&-\frac{z}{sH(M_{\chi_1})}\left[\left(\frac{Y_{\chi_1}}{Y_{\chi_1}^{eq}} - 1\right) \left(\gamma_{D}^{(1)}\right) + \left[\left(\frac{Y_{\chi_1}}{Y_{\chi_1}^{eq}}\right)^2 - 1\right]\gamma_{z'}\right]\\
    \frac{dY_{\chi_2}}{dz}&=&-\frac{z}{sH(M_{\chi_1})}\left[\left(\frac{Y_{\chi_2}}{Y_{\chi_2}^{eq}} - 1\right) \left(\gamma_{D}^{(2)}\right) + \left[\left(\frac{Y_{\chi_2}}{Y_{\chi_2}^{eq}}\right)^2 - 1\right]\gamma_{z'}\right]\\
    \frac{Y_{B-L}}{dz}&=&-\frac{z}{sH(M_{\chi_1)}}\Bigg[\sum_{j=1}^{2}\left[0.5\frac{Y_{B-L}}{Y_{l}^{eq}} + \epsilon_{j}\left(\frac{Y_{\chi_1}}{Y_{\chi_1}^{eq}} - 1\right)\right]\gamma_{D_{j}}\Bigg]
\end{eqnarray}
where $z=\frac{M_{\chi_1}}{T}$ and $H(M_{\chi_1})=1.66\sqrt{g_{eff}}\frac{T^2}{M_{Pl}}|_{T=M_{\chi_1}}$, $M_{Pl}=1.22\times10^{19}\quad GeV$ is the Planck Mass, $g_{eff}\approx110$ is total degrees of freedom and $s$ is the entropy density. The right-hand side of the above equations represents the interaction in which the $\chi_i$ particle takes part. The equilibrium comoving densities are given by the following relations \cite{Behera:2020sfe,Davidson:2008bu}:
\begin{eqnarray}
    Y_{\chi_i}^{eq} &=& \frac{45g_{\chi_i}}{4\pi^{4}g_{eff}}z^{2}K_{2}(z)\\
    Y_{l}^{eq}&=&\frac{3}{4}\frac{45g_{l}\zeta(3)}{2\pi^{4}g_{eff}}
\end{eqnarray}
Here, $g_{\chi_i} = 2$ and $g_{l} = 2$ are degrees of freedom of the right-handed and lepton superfield, respectively.
In the numerical analysis, we have omitted the scattering processes mediated by the Higgs boson and the right-handed neutrinos as their effects are negligible compared to $\gamma_{D}$ and $\gamma_{Z'}$ (See Appendix:\ref{sec: Append C}).
\par
 
\begin{figure}
        \centering
        \includegraphics[scale=0.35]{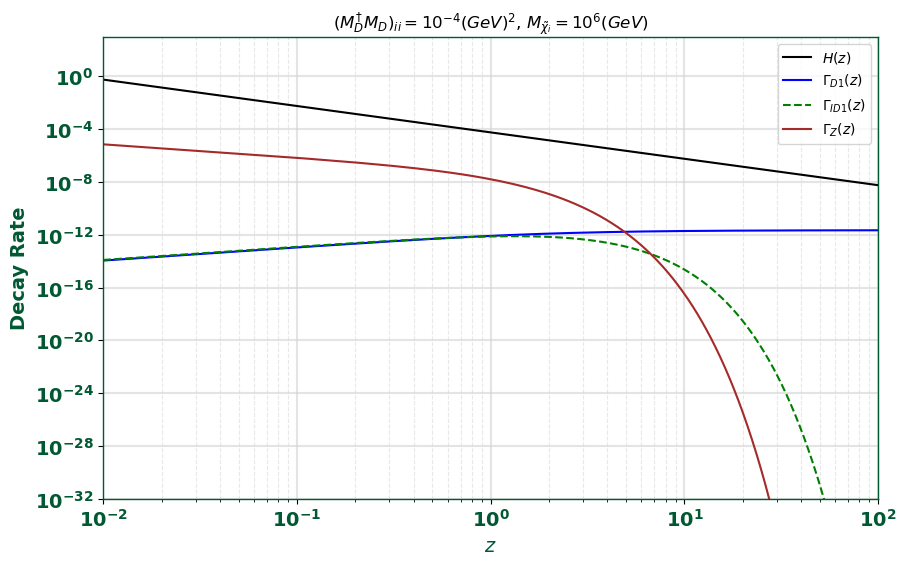}
        \includegraphics[scale=0.33]{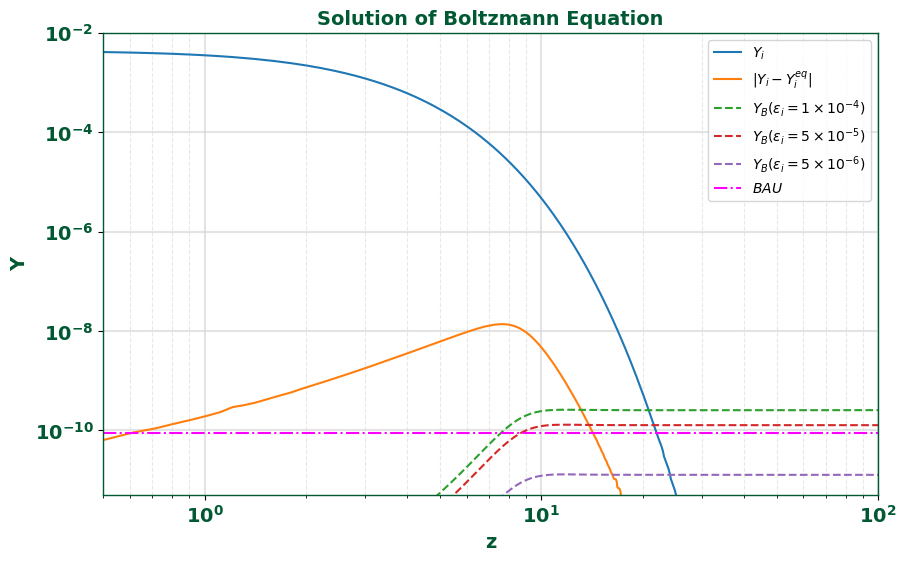}
        \includegraphics[scale=0.33]{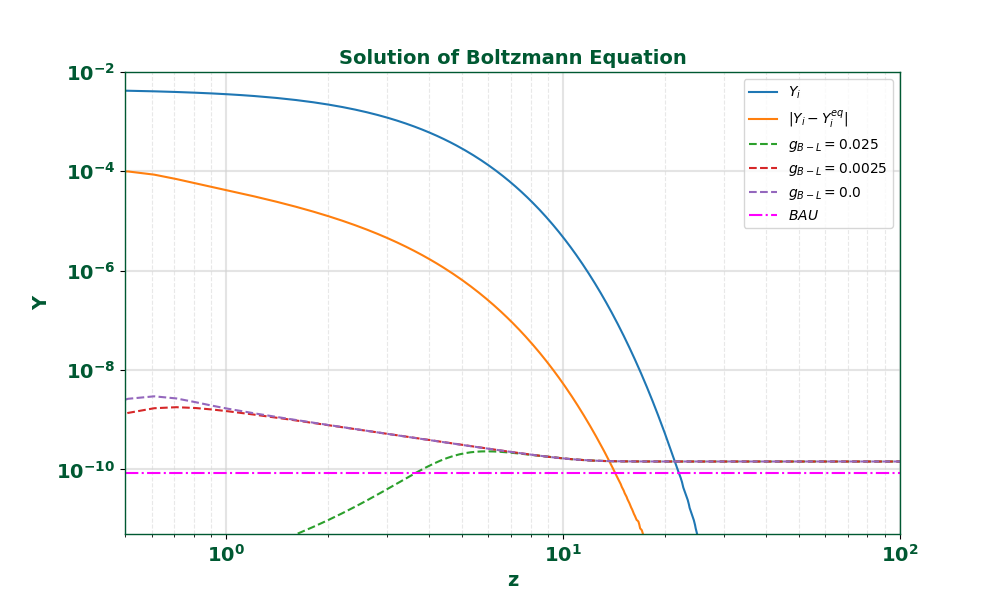}
        \caption{(\textit{Top-left}) The decay rate, inverse decay rate, and scattering rate comparison with Hubble expansion rate. (\textit{Top-right}) For different values of the CP-parameter $\epsilon_i$, the baryon asymmetry is shown against  $z=\frac{M_{\chi_{1}}}{T}$. (\textit{Bottom}) The baryon asymmetry for different values of gauge coupling $g_{B-L}$.} 
       \label{fig:fig-BAU} 
\end{figure}

 In the \textit{top-left} panel of Fig.\ref{fig:fig-BAU} we have shown interaction rates with Hubble expansion. The solid blue line for the decay process, and the dashed green line for the inverse decay process of $\chi_{1}$. Similar interaction rate lines are obtained in the case of $\chi_{2}$. The $Z'$ mediated interaction rate is shown by the solid red line. The generation of baryon asymmetry for different values of CP asymmetry $\epsilon_{i}$ is shown in the \textit{top-right} panel of Fig.\ref{fig:fig-BAU}. Once the out-of-equilibrium condition is achieved, the decay (blue curve) proceeds slowly. As a result, $Y_i$ no longer follows $Y_{i}^{eq}$ (yellow curve) and lepton asymmetry is generated (dashed curves). The obtained lepton asymmetry converted into baryon asymmetry via the sphaleron process and is given by Eq.\eqref{Eq of YB}. We have taken $m_{Z'}=4$ TeV (The mass of $Z'$ boson)\cite{ParticleDataGroup:2024cfk}, $(M_{D}^{\dagger}M_{D})_{ii}=10^{-4} (GeV^2)$ and $M_{\chi_{i}}=10^6 (GeV)$ in the calculations. For this value of $Z'$ mass, we obtain different values $g_{B-L}$ in the range $(0.014-0.000014)$ for the VEV range $v_{\phi} \in [10^5, 10^8]$ GeV, which satisfy the neutrino oscillation data. In the \textit{bottom} panel of Fig.\ref{fig:fig-BAU} we have shown the obtained baryon asymmetry for $g_{B-L} = 0.025$ (dashed green line) and $g_{B-L} = 0.0025$ (dashed red line). We have also shown for $g_{B-L} = 0$ (dashed purple line), which corresponds to the absence of $Z'$ mediated interaction. It can be seen that a larger $g_{B-L}$ tries to delay the generation of the baryon asymmetry. Because a larger $g_{B-L}$ means larger scattering and brings the $\chi_{i}$ into thermal equilibrium efficiently. Since baryon asymmetry can only arise from out-of-equilibrium decays, a larger value of \( g_{B-L} \), which keeps the system closer to equilibrium, is expected to result in a smaller generated baryon asymmetry. However, the final asymmetry turns out to be independent of \( g_{B-L} \), because although the generation of asymmetry is delayed, the \( \chi_i \) particles still decay efficiently. As a result, the decay process experiences minimal washout, allowing the full asymmetry to be preserved. The horizontal dashed pink line in Fig.\ref{fig:fig-BAU} represents the current bound on baryon asymmetry $Y_{B} \sim  8.6*10^{-11}$. 

%%%%%%%%%%%----------Conclusion------------------------%%%%%%%%%%%%%%%%
\section{Conclusion}\label{sec5}

We study the minimal inverse seesaw in our work using $A_4$ modular symmetry. Our model simultaneously addressed neutrino oscillation parameters and matter-antimatter asymmetry. Certain regions of \( \text{Re}[\tau] \) and \( \text{Im}[\tau] \) are ruled out by our model. The $\text{Re}[\tau]$ value in the region $(-0.11,0.98)$ and $\text{Im}[\tau]$ value in the region $(0.85,0.98)$ are excluded. All mixing angles are found within the $3\sigma$ bound, and our model imposes a lower bound of 0.44 on the atmospheric mixing angle \( \sin^{2}\theta_{23} \). There is a linear relationship between the mixing angles $\sin^{2}\theta_{23}$ and $\sin^{2}\theta_{12}$. The angle $\sin^{2}\theta_{12}$ decreases with increasing $\sin^{2}\theta_{23}$ within the $3\sigma$ region of both mixing angles. The sum of the neutrino mass for the model is found to be in the range $\sum m_{\nu} \in [0.0580-0.0605]$ eV in NO. Our model also predicts a strong correlation of atmospheric mixing angle($\sin^{2}{\theta_{23}}$) with Jarkslog invariant($J_{CP}$) and Dirac CP phase ($\delta_{CP}$). The parameter space consistent with neutrino oscillation data also satisfies the experimental constraints from LFV decays. We have also studied neutrinoless double beta decay, and the effective neutrino mass value is found well below the current experiment sensitivity limit. Using the same parameter space that explains neutrino oscillation data, we have also investigated the matter-antimatter asymmetry and found that it is consistent with the current observed value.

%%%%%%%%%%%%%%%%%%    Acknowledgement %%%%%%%%%%%%%%%%%%%%%%%%%%%%

\section*{Acknowledgement}
GP would like to acknowledge CSIR-HRDG for the financial support received in the form of JRF fellowship (09/0796(16046)/2022-EMR-I). 

%%%%%%%%%%%%%%%%%%%%%     Appendix     %%%%%%%%%%%%%%%%%%%%%%%%
%%%%%%%%%%%      Appendix A         %%%%%%%%%%%%%%%%%%%%%%%%%%%%
\appendix

\renewcommand{\theequation}{A\arabic{equation}} % Prefix equations with "A"

\setcounter{equation}{0} % Reset equation counter
%\begin{center}
\section{\(A_4\) Group}
$A_4$ is the symmetry group of the tetrahedron and the group of even permutations of four objects. It therefore has $4!/2 = 12$ elements. It can be seen that all twelve elements can be obtained by repeatedly multiplying the two generators, $S = (14)(23)$ and $T = (123)$. These satisfy the relations:
\begin{equation*}
    S^2=(ST)^3=T^3=1.    
\end{equation*}
 $A_4$ has four inequivalent representations: three of dimension one, 1, $1'$ and $1''$ and one of dimension 3. The product rule for $A_4$ group  is given as follow:
 \begin{equation}  
\begin{split}
   & 1\otimes 1=1\\
   & 1'\otimes 1''=1\\
   & 1'\otimes 1'=1''\\
   & 1^{(')('')}\otimes 3=3\\
   & 3\otimes 3= 1\oplus 1'\oplus 1''\oplus 3_{S}\oplus 3_{A}
\end{split}    
\end{equation}
where, $3_{S(A)}$ represents symmetric (anti-symmetric) combination. There are two basis for triplet multiplication: Ma-Rajasekaran basis and Altarelli-Feruglio basis.
 In the Altarelli-Feruglio (AF) basis the generator $T$ is diagonal.
 \begin{equation}
 \begin{split}
     T = \begin{pmatrix}
         1 & 0 & 0 \\
         0 & \omega & 0 \\
         0 & 0 & \omega^2 \\
     \end{pmatrix}, \,
     S = \frac{1}{3} \begin{pmatrix}
         -1 & 2 & 2 \\
         2 & -1 & 2 \\
         2 & 2 & -1
     \end{pmatrix} .
 \end{split}
 \end{equation}
 Here, $\omega$ is the cubic root of unity.\\
 The multiplication rule for $3\otimes 3$ is given for triplets $a=(a_1, a_2, a_3)$ and $b=(b_1, b_2, b_3)$ in the AF basis as follows:
  \begin{equation}
 \begin{split}
 & 1_{AF}: (ab)_1 = a_1 b_1 + a_3 b_2 + a_2 b_3,\\
 & 1'_{AF}: (ab)_{1'} = a_1 b_2 + a_2 b_1 + a_3 b_3,\\
 & 1''_{AF}: (ab)_{1''} = a_1 b_3 + a_3 b_1 + a_2 b_2,\\
 & 3_{AF}^s: \frac{1}{\sqrt{3}}(2a_1 b_1 - a_2 b_3 - a_3 b_2, 2a_3 b_3 - a_1 b_2 - a_2 b_1, 2a_2 b_2 - a_3 b_1 - a_1 b_3)\\
 & 3_{AF}^A: (a_2 b_3 - a_3 b_2, a_1 b_2 - a_2 b_1, a_3 b_1 - a_1 b_3)
 \end{split}    
 \end{equation}
%%%%%%%%%%%%%%%%%     Appendix B %%%%%%%%%%%%%%%%%%%%%%%%%%
\renewcommand{\theequation}{B\arabic{equation}} % Prefix equations with "B"
\setcounter{equation}{0} % Reset equation counter
\section{ Modular forms of Yukawa couplings}
\label{sec: Append B}

$\bar{\Gamma}$ is the modular group that attains a linear fractional transformation $\gamma$ which acts on modulus $\tau$ linked to the upper-half complex plane whose transformation is given by:
\begin{equation}
\gamma \longrightarrow \gamma \tau = \frac{a\tau + b}{c\tau + d},
\end{equation}
where $a, b, c, d \in  \mathbb{Z}$ and $ad-bc = 1$, $Im[\tau]>0$, where it is isomorphic to the transformation $PSL(2, \mathbb{Z}) = SL(2, \mathbb{Z})/\{I, -I\}$. The S and T transformation helps in generating the modular transformation defined by:
\begin{equation}
    S: \tau \longrightarrow -\frac{1}{\tau}, \hspace{1.5cm} T: \tau \longrightarrow \tau + 1,
\end{equation}
and hence the algebraic relations so satisfied are as follows,
\begin{equation}
    S^2 = \mathbb{I}, \hspace{2.5cm} (ST)^3 = \mathbb{I}.
\end{equation}
Here, series of groups are introduced, $\Gamma(N) (N = 1, 2, 3, .....)$ and defined as
\begin{equation}
    \Gamma(N) =\Bigg \{\begin{pmatrix}
        a & b \\
        c & d
    \end{pmatrix} \in SL(2,\mathbb{Z}), \begin{pmatrix}
        a & b \\
        c & d 
    \end{pmatrix} = \begin{pmatrix}
        1 & 0 \\
        0 & 1
    \end{pmatrix}(\text{mod} N)\Bigg\}
\end{equation}
Definition of $\bar{\Gamma}(2) \equiv \Gamma(2)/\{I, -I\}$ for N=2. Since $-I$ is not associated with $\Gamma(N)$ for $N>2$ case, one can have $\bar{\Gamma}(N)=\Gamma(N)$, which are infinite normal subgroups of $\bar{\Gamma}$ known as principal congruence subgroups. Quotient groups come from the finite modular group, defined as $\Gamma_N = \bar{\Gamma}/\bar{\Gamma}(N)$. The
imposition of $T^N = \mathbb{I}$ is done for these finite groups $\Gamma_N$. Thus, the groups $\Gamma_N (N = 2, 3, 4, 5)$ are isomorphic to $S_3$, $A_4$, $S_4$ and $A_5$, respectively. N level modular forms are holomorphic
functions $f(\tau)$ which are transformed under the influence of $\Gamma(N)$ as follows:
\begin{equation}
    f(\gamma\tau)=(c\tau + d)^{k}f(\tau), \hspace{1.5cm} \gamma \in \Gamma(N)
\end{equation}
where k is the modular weight. 

%%%%%%%%%%%%%%%%%%%%%%%%%%%%%%%%%%%%%%%%%%%%%%%%%%%%%%%%%%%%%%%%

 A field $\phi^{(I)}$ transforms under the modular transformation as:
 \begin{equation}
     \phi^{(I)} \rightarrow (c\tau + d)^{-k_I}\rho^{I}(\gamma)\phi^{(I)}
 \end{equation}
 where $-k_I$ represents the modular weight and $\rho^{(I)}(\gamma)$ signifies an unitary representation matrix of $\gamma \in \Gamma(2)$.

%%%%%%%%%%%%%%%%%%%%%%%%%%%%%%%%%%%%%%%%%%%%%%%%%%%%%%%%%%%%%%%%%%

The scalar field's kinetic term is as follows:
\begin{equation}
    \sum_{I} \frac{|{\partial_{\mu}\phi^{(I)}}|^2}{(-i\tau + i\bar{\tau})^{k_I}}
\end{equation}

%%%%%%%%%%%%%%%%%%%%%%%%%%%%%%%%%%%%%%%%%%%%%%%%%%%%%%%%%%%%%%%%%%%%

The modular forms of the Yukawa coupling $Y = (y_1, y_2, y_3)$ with weight 2, which transforms as a triplet of $A_4$ can be expressed in terms of Dedekind eta-function $\eta(\tau)$ and its derivative:
\begin{equation}
\begin{split}
& y_1(\tau) = \frac{i}{2\pi}\bigg(\frac{\eta'(\tau/3)}{\eta(\tau/3)} + \frac{\eta'((\tau + 1)/3)}{\eta((\tau+1)/3)} + \frac{\eta'((\tau+2)/3)}{\eta((\tau+2)/3)} - \frac{27\eta'(3\tau)}{\eta(3\tau)}\bigg), \\  
& y_2(\tau) = \frac{-i}{\pi}\bigg(\frac{\eta'(\tau/3)}{\eta(\tau/3)} + \omega^2 \frac{\eta'((\tau + 1)/3)}{\eta((\tau+1)/3)} + \omega \frac{\eta'((\tau+2)/3)}{\eta((\tau+2)/3)}\bigg), \\
& y_3(\tau) = \frac{-i}{\pi}\bigg(\frac{\eta'(\tau/3)}{\eta(\tau/3)} + \omega \frac{\eta'((\tau + 1)/3)}{\eta((\tau+1)/3)} + \omega^2 \frac{\eta'((\tau+2)/3)}{\eta((\tau+2)/3)}\bigg), \\
\end{split}    
\end{equation}
The Dedekind eta-function $\eta(\tau)$ is given by:
\begin{equation}
    \eta(\tau) = q^{1/24} \prod_{n=1}^{\infty}(1 - q^n), \hspace{1.5cm} q \equiv e^{i2\pi\tau}
\end{equation}
In the form of q-expansion, the modular Yukawa can be expressed as:
\begin{equation}
\begin{split}
 &   y_1(\tau) = 1 + 12q + 36q^2 + 12q^3 + ............., \\
  &  y_2(\tau) = -6q^{1/3}(1 + 7q + 8q^2 + .............), \\
   & y_3(\tau) = -18q^{2/3}(1 + 2q + 5q^2 + .............).
\end{split}    
\end{equation}
From the q-expansion, we have the following constraint for modular Yukawa couplings:
\begin{equation}
    y_2^{2} + 2 y_1 y_3 = 0
\end{equation}
Higher modular weight Yukawa couplings can be constructed from weight 2 Yukawa $\mathbf{(Y^{(2)})}$ using the $A_4$ multiplication rule. For modular weight k = 4, we have the following five modular forms:
\begin{equation}
\begin{split}
  &  Y_{3}^{(4)} = (y_1^2 - y_2 y_3, y_{3}^{2}-y_1 y_2, y_2^2 - y_1 y_3),\\
   & Y_{1}^{(4)} = y_1^2 + 2 y_2 y_3,\\
   & Y_{1'}^{(4)} = y_3^2 + 2 y_1 y_2
\end{split}    
\end{equation}
For modular weight k = 6, there are seven modular forms:
\begin{equation}
\begin{split}
& Y_1^{(6)} = y_1^3 + y_2^3 + y_3^3 - 3y_1y_2y_3,\\
& Y_{3a}^{(6)}=(y_{31},y_{32},y_{33}) = (y_1^3 + 2y_1y_2y_3, y_1^2y_2 + 2 y_2^2y_3, y_1^2y_3 + 2y_3^2y_2),\\
& Y_{3b}^{(6)} = (y_3^3 + 2y_1y_2y_3, y_3^2y_1 + 2 y_1^2y_2, y_3^2y_2 + 2y_2^2y_1)
\end{split}
\label{eq:weight6}
\end{equation} 
For modular weight k=8, there are nine modular forms:
\begin{equation}
\begin{split}
& Y_1^{(8)} = (Y_1^{2} + 2Y_2Y_3)^2, \quad Y_{1'}^{(8)} = (Y_1^{2} + 2Y_2Y_3)(Y_3^{2} + 2Y_1Y_2), \quad Y_{1''}^{(8)} = (Y_3^{2} + 2Y_1Y_2)^2,\\
& Y_{3a}^{(8)} = (Y_1^{2} + 2Y_2Y_3)\begin{pmatrix}
    Y_1^{2} - Y_2Y_3\\
    Y_3^{2} - Y_1Y_2\\
    Y_2^{2} - Y_1Y_3
\end{pmatrix}, \quad Y_{3b}^{(8)} = (Y_3^{2} + 2Y_1Y_2)\begin{pmatrix}
    Y_2^{2} - Y_1Y_3\\
    Y_1^{2} - Y_2Y_3\\
    Y_3^{2} - Y_1Y_2
\end{pmatrix}
\end{split}
\end{equation}
For modular weight k=10, there are eleven modular forms:
\begin{equation}
\begin{split}
& Y_1^{(10)} = (Y_1^{2} + 2Y_2Y_3)(Y_1^{3} + Y_2^{3} + Y_3^{3} - 3Y_1Y_2Y_3),\\
& Y_{1'}^{(10)} = (Y_3^{2} + 2Y_1Y_2)(Y_1^{3} + Y_2^{3} + Y_3^{3} - 3Y_1Y_2Y_3),\\
& Y_{3a}^{(10)} = (Y_1^{2} + 2Y_2Y_3)^2\begin{pmatrix}
    Y_1\\
    Y_2\\
    Y_3
\end{pmatrix}, \quad Y_{3b}^{(10)} = (Y_3^{2} + 2Y_1Y_2)^2\begin{pmatrix}
    Y_2\\
    Y_3\\
    Y_1
\end{pmatrix},\\
& Y_{3c}^{(10)} = (Y_1^{2} + 2Y_2Y_3)(Y_3^{2} + 2Y_1Y_2)\begin{pmatrix}
    Y_3\\
    Y_1\\
    Y_2
\end{pmatrix}.
\end{split}
\end{equation}
For modular weight k=12, there are thirteen modular forms:
\begin{equation}
\begin{split}
& Y_{1a}^{(12)} = (Y_1^{2} + 2Y_2Y_3)^{3}, \quad Y_{1b}^{(12)} = (Y_3^{2} + 2Y_1Y_2)^{3},\\
& Y_{1'}^{(12)} = (Y_1^{2} + 2Y_2Y_3)^{2}(Y_3^{2} + 2Y_1Y_2), \quad Y_{1''}^{(12)} = (Y_1^{2} + 2Y_2Y_3)(Y_3^{2} + 2Y_1Y_2)^{2},\\
& Y_{3a}^{(12)} = 2(Y_1^{2} + 2Y_2Y_3)^{2}\begin{pmatrix}
    Y_1^{2} - Y_2Y_3\\
    Y_3^{2} - Y_1Y_2\\
    Y_2^{2} - Y_1Y_3
\end{pmatrix}, \quad Y_{3b}^{(12)} = -2(Y_3^{2} + 2Y_1Y_2)^{2}\begin{pmatrix}
    Y_3^{2} - Y_1Y_2\\
    Y_2^{2} - Y_1Y_3\\
    Y_1^{2} - Y_2Y_3
\end{pmatrix},\\
& Y_{3c}^{(12)} = (Y_1^{2} + 2Y_2Y_3)(Y_3^{2} + 2Y_1Y_2)\begin{pmatrix}
    Y_2^{2} - Y_1Y_3\\
    Y_1^{2} - Y_2Y_3\\
    Y_3^{2} - Y_1Y_2
\end{pmatrix}.
\end{split}
\label{eq:weight12}
\end{equation}

%%%%%%%%%%%%%%%%%%%%%%%%%%%%%%%%%%%%%%%%%%%%%%%%%%%%%%%%%%%%%%%
%%%%%%%%%%%%%%% Appendix C %%%%%%%%%%%%%%%%%%%%%%%%
\renewcommand{\theequation}{C\arabic{equation}} % Prefix equations with "B"
\setcounter{equation}{0} % Reset equation counter
\section{ Decay and scattering rates:}
\label{sec: Append C}

The relation for decay and two-particle scattering processes is given in the following:\\
\textit{Dacay rates:}\\
The decay rates are given by \cite{Plumacher:1996kc}:
\begin{equation}
    \gamma_{D} := \gamma^{eq}({\chi}\longrightarrow i + j + .......) = sY_{{\chi}}^{eq}\Gamma_{D}
\end{equation}
where, $\Gamma_D = \Gamma_{{\chi}}\frac{K_{1}(z)}{K_{2}(z)}$ and here $K_{1}(z)$ and $k_{2}(z)$ are modified Bessel functions. Also, the decay width of ${\chi_i}$ at tree level is given by $\Gamma_{{\chi_i}}=\frac{\alpha}{\sin^{2}{\theta_W}}\frac{M_{{\chi_i}}}{4}\frac{(M_{D}^{\dagger}M_D)_{ii}}{M_{W}^2}$ with $\alpha, \theta_W, M_W$ being fine structure constant, Weinberg angle and W-boson mass respectively.\\
\textit{Two body scattering rates:}\\
In the case of two-body scattering, the reaction density is given by \cite{Plumacher:1996kc}:
\begin{equation}
    \gamma^{eq}({\chi} + a \longleftrightarrow i + j + ......) = \frac{T}{64\pi^{4}}\int_{s_{min}}^{\infty}ds\hat{\sigma}(s)\sqrt{s}K_{1}\left(\frac{\sqrt{s}}{T}\right)
\end{equation}
where $s_{min}=Max[(M_{{\chi}}+M_a)^{2},(M_i + M_j)^2]$, $s$ is the center of mass energy and the reduced cross-section $\hat{\sigma}(s)$ for the process is related to the usual total cross-section $\sigma(s)$ as follows:
\begin{equation}
    \hat{\sigma}(s) = \frac{8}{s}\left[(p_{{\chi}}.p_{a})^{2} - M_{{\chi}}^{2}M_{a}^{2}\right]\sigma(s)
\end{equation}
where $p_{\chi}$ and $M_{\chi}$ are the three momentum and mass of $\chi$ particle respectively.\\
The dominating contribution comes from the $Z'-$ exchange pair creation $s-$ channel process. The reduced cross-section for this process is given as follows \cite{Plumacher:1996kc}:
\begin{equation}
    \hat{\sigma}_{Z'}(s) = \frac{4225\pi\alpha^2_{B-L}}{216\cos^{4}{\theta_W}}\frac{\sqrt{x}}{(x-y)^2 + yb}(x-4a_1)^{3/2}.
\end{equation}
where  $y=\frac{M_{Z'}^2}{M_{{\chi_1}}^2}$, $x = \frac{s}{M_{\chi_1}^2}$, $b=\left(\frac{\Gamma_{Z'}}{M_{{\chi_1}}}\right)^2$, $\alpha_{B-L} = \frac{g^2_{B-L}}{4\pi}$ and the decay width of $Z'$ is given by the relation $\Gamma_{Z'}=\frac{\alpha_{B-L} M_{Z'}}{\cos^{2}{\theta_W}}\left[\frac{25}{18}\sum_{i}\left(\frac{y-4a_i}{4y}\right)^{3/2}\theta(y-4a_i) + \frac{169}{144}\right]$.

%%%%%%%%%%%%%%%%%%%%%%%%%%%%%%%%%%%%%%%%%%%%%%%%%%%%%%%%%%%%%%%%%

\bibliographystyle{JHEP}
%\bibstyle{apsrev}
\bibliography{reference}
\end{document}